\documentclass[10pt, conference]{IEEEtran}
\usepackage{cite}
\usepackage{amsmath,amssymb,amsfonts}
\usepackage{algorithmic}
\usepackage{graphicx}
\usepackage{textcomp}
\usepackage{xcolor}
\usepackage[hyphens]{url}
\usepackage{array}
\usepackage{tabularx}
\usepackage{placeins}
\usepackage{fancyhdr}
\usepackage{afterpage}
\def\BibTeX{{\rm B\kern-.05em{\sc i\kern-.025em b}\kern-.08em
    T\kern-.1667em\lower.7ex\hbox{E}\kern-.125emX}}

\makeatletter
\def\hlinewd#1{%
\noalign{\ifnum0=`}\fi\hrule \@height #1 %
\makeatother
\futurelet\reserved@a\@xhline}
\usepackage{booktabs}
\usepackage[linesnumbered,ruled,vlined]{algorithm2e}
\usepackage{footnote}
\usepackage{hyperref}
\usepackage{float}
\usepackage{dblfloatfix}
\definecolor{samblue}{RGB}{20,40,160}
\definecolor{lightgreen}{RGB}{241,248,236}
\definecolor{hlg}{RGB}{225,255,235}
\definecolor{lightblack}{RGB}{63,63,63}
\definecolor{lightgray1}{RGB}{245,245,245}
\definecolor{lightgray2}{RGB}{216,216,216}
\definecolor{figblue}{RGB}{20,40,160}
\definecolor{figred}{RGB}{255,0,0}
\usepackage{tikz}
\hypersetup{colorlinks=true, citecolor=black, linkcolor=black, urlcolor=black}
\DeclareRobustCommand\bcirc[1]{\tikz[baseline=(char.base)]{\node[shape=circle,draw,inner sep=0pt,fill=black, text=white] (char) {#1};}}

\renewcommand\IEEEkeywordsname{Keywords}
\renewcommand\#{\protect\scalebox{1}{\protect\raisebox{0ex}{\char"0023}}}
\usepackage{multirow}
\usepackage{colortbl}
\usepackage{amssymb}
\usepackage{pifont}

\usepackage{soul}
\soulregister\ref7
\usepackage{authblk}
\usepackage{adjustbox}
\usepackage{threeparttable}
\usepackage[skins]{tcolorbox}

\usepackage{enumitem}
\usepackage{titlesec}
\usepackage{fix-cm}

\title{DSAC: Low-Cost RowHammer Mitigation Using In-DRAM Stochastic and Approximate Counting Algorithm}

\author[]{Seungki Hong}
\author[]{Dongha Kim}
\author[]{Jaehyung Lee}
\author[]{Reum Oh}
\author[]{Changsik Yoo}
\author[]{Sangjoon Hwang}
\author[]{Jooyoung Lee}

\affil[]{DRAM Design Team, Memory Division, Samsung Electronics}

\begin{document}
\maketitle
\thispagestyle{plain}
\pagestyle{plain}

\begingroup
\renewcommand\thefootnote{}
\footnotetext{\scriptsize Version 2 was submitted to HPCA ’23 but was rejected. A revised version with additional data was later accepted to ISCA ’24; however, the author subsequently requested its withdrawal due to a potential issue. This version is based on the original revision, excluding the additional data, and is provided solely to improve clarity and figure rendering compared to Version 2.

Additionally, the authors’ affiliations have changed since then. Therefore, this version does not represent the current work of Samsung, and no further revisions will be made. Nevertheless, the original authors and affiliations are retained, as the work was completed during the period in which Version 2 was developed.}
\endgroup


\begin{abstract}
This paper provides the fundamental mechanisms of two types of row activation-induced bit flips and proposes in-DRAM protection techniques. \textit{RowBleed} occurs when a victim row experiences charge leakage due to transistor's threshold voltage lowering induced by long activation of a neighboring aggressor row. Therefore, this paper proposes \textit{Time-Weighted Counting} for RowBleed mitigation, which assigns greater counter weights to rows that are activated for longer durations.

On the other hand, \textit{RowHammer} occurs when a victim row experiences electron injection due to frequent activation of a neighboring aggressor row. Similarly, Extended RowHammer, the phenomenon where victim rows are two rows beyond aggressor rows, is also caused by electron injection due to frequent activation of a neighboring aggressor row. Consequently, accurate detection of aggressor rows is crucial. Therefore, this paper proposes RowHammer mitigation algorithm named \textit{DSAC} (in-DRAM Stochastic and Approximate Counting algorithm), which utilizes a replacement probability that adjusts based on the count of the old row.

This paper introduces a RowHammer protection index called Maximum Disturbance, which measures the maximum accumulated number of row activations within an observation period. The experimental results demonstrate that DSAC can achieve 133x lower Maximum Disturbance than the state-of-the-art counter-based algorithm.
\end{abstract}

\section{Introduction}
DRAM is a type of volatile memory that stores data in cells consisting of one capacitor and one transistor. However, DRAM manufacturers have scaled down the size of these cells to achieve a lower cost per bit, which has increased electromagnetic crosstalk between cells. This crosstalk can negatively affect row activation, which must occur for the system to read or write data, causing row activation-induced bit-flips. In 2012, Intel claimed that Samsung's commodity DRAM was vulnerable to frequent row activations \cite{dramsec}. Since then, the first academic paper \cite{kim2014flipping} discussed this phenomenon, and many subsequent studies \cite{cojocar2019exploiting, gruss2018another, gruss2016RowHammer, kurmus2017random, lipp2020nethammer, seaborn2015exploiting, tatar2018thRowHammer, van2016drammer, van2018guardion, walker2021dram, xiao2016one, yang2016suppression} have explored RowHammer. Various Target-Row-Refresh (TRR) algorithms, which refresh victim rows, have also been proposed to mitigate the disturbance caused by RowHammer \cite{you2019mrloc, lee2019twice, kang2020cat, kim2014architectural, park2020graphene, seyedzadeh2016counter, son2017making}. However, a recent study \cite{frigo2020trrespass} has shown that RowHammer is still a problem, which can be exacerbated by the shrinking distance between DRAM cells, decreasing the RowHammer threshold (RH$_{\text{TH}}$), which is the number of activations required for RowHammer-induced bit-flips.

The motivation of this paper is to present an industrial perspective on row activation-induced bit-flips, providing crucial insights for future research to enhance DRAM security against such vulnerabilities. Firstly, this paper investigates the fundamental mechanisms of row activation-induced bit-flips at the cell level. Secondly, the paper elucidates the limitations of state-of-the-art counter-based algorithms in effectively detecting RowHammer, due to DRAM manufacturers' focus on low area cost.

In Section \ref{sec:2}, this paper discusses the fundamental mechanisms of row activation-induced bit-flips. This paper categorizes row activation-induced bit-flips into two types: RowBleed and RowHammer. RowBleed occurs when a victim row experiences charge leakage due to transistor's threshold voltage lowering induced by long activation of a neighboring aggressor row. On the other hand, RowHammer occurs when a victim row experiences electron injection due to frequent activation of a neighboring aggressor row. Similarly, Extended RowHammer, the phenomenon where victim rows are two rows beyond aggressor rows, is also caused by electron injection due to frequent activation of a neighboring aggressor row. Additionally, this paper demonstrates that the critical factor in RowHammer is not only the number of row activations but also row precharge-to-activation time.

In Section \ref{sec:4}, this paper proposes Time-Weighted Counting for RowBleed mitigation, which assigns greater counter weights to rows that are activated for longer durations. This paper also proposes RowHammer mitigation algorithm named DSAC, which can filter out decoy-rows. Decoy-rows are rows whose number of accesses does not exceed the number of RowHammer accesses. However, the state-of-the-art counter-based algorithms cannot filter out decoy-rows due to their constant replacement probability. Consequently, decoy-rows can replace RowHammer entries in a count table, thereby diminishing TRR chances for victim rows. DSAC addresses this issue by utilizing a replacement probability that adjusts based on the count of the old row. The core concept is that, on average, only rows with counts exceeding a minimum count in a count table can replace rows with that minimum count in a count table.

This paper makes the following key contributions:
\begin{itemize}[leftmargin=*,align=left, noitemsep, topsep=0pt, parsep=0pt]
\item This paper provides a comprehensive understanding of two types of row activation-induced bit flips: RowBleed, caused by charge leakage, and RowHammer, caused by electron injection.
\item This paper proposes Time-Weighted Counting for RowBleed mitigation, which assigns greater counter weights to rows that are activated for longer durations.
\item This paper also proposes RowHammer mitigation algorithm named DSAC, which can filter out decoy-rows by utilizing a replacement probability that adjusts based on the count of the old row.
\end{itemize}

\section{Background}
In this section, the necessary background on DRAM organization and operation is described.

\begin{figure}[h]
    \centering
    \includegraphics[width=0.9\columnwidth]{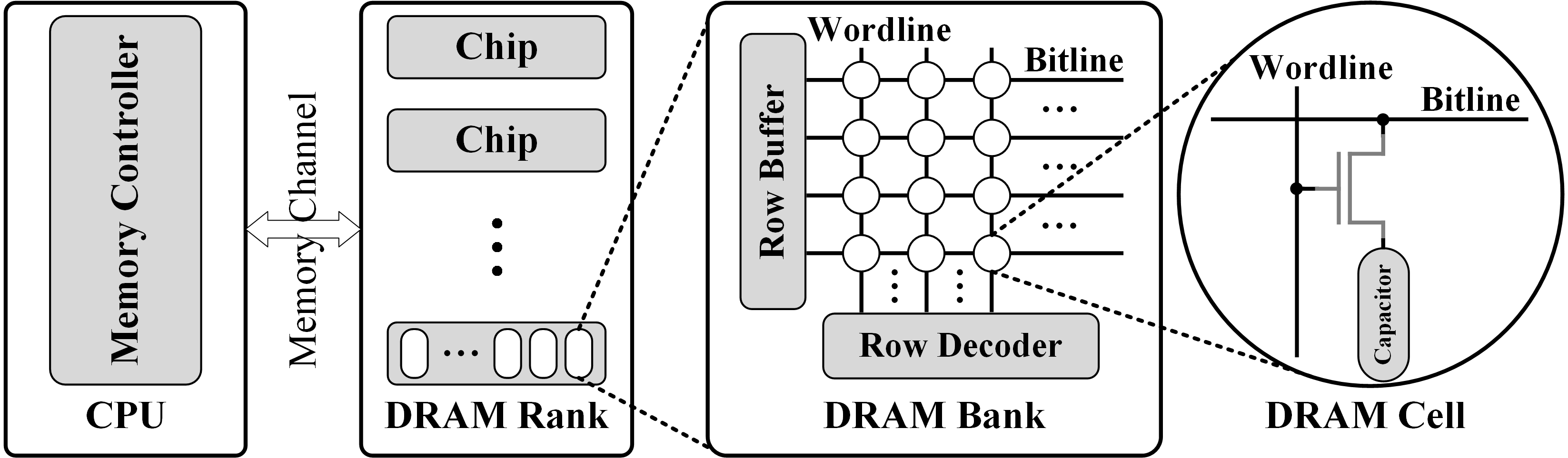}
    \vspace{-0.5em}
    \caption{Architecture of Typical DRAM-Based System}
    \label{fig:cell}
\end{figure}

Figure \ref{fig:cell} represents a typical computer architecture with emphasis on DRAM. A bank comprises many subarrays and each subarray contains a two-dimensional array of cells arranged in rows and columns. When accessing each cell, the row decoder first decodes incoming row address to open the row by driving the corresponding wordline. To write or read the stored data in cells, each row needs to be in an active state. The row must be precharged before further accesses can be made to other rows of the same bank. A DRAM cell consists of a single capacitor and a transistor. Since the transistor leaks its current, the capacitor leaks its charge over time. To prevent this cell data loss, DRAM cells need to be periodically refreshed. Thus, memory controller periodically issues refresh commands to DRAM.

This paper sets the baseline parameters in Table \ref{tab:base} adhering to memory standard specifications \cite{specification2014mjedec, specification2019jedec, specification2014jedec, specification2020jedec, specification2021gjedec, specification2021hjedec}.

\begin{table}[h]
\vspace{-1ex}
\centering
\caption{Baseline Parameters}
\vspace{-1ex}
\label{tab:base}
    \begin{adjustbox}{width=\columnwidth}
        \begin{threeparttable}
        \begin{tabular}{lll}
        \hlinewd{1.2pt}
        {Parameter} & {Description} & {Value} \\ \hline
        {tREFI} & Ref. CMD Interval & {15.625us (MR4 4x)} \\ 
        {tREFW} & 8K Ref. CMD Window ($\text{tREFI}\times{\text{8K}}$) & {128ms (MR4 4x)} \\
        {tRFC} & Ref. Operation Time & {280ns (8Gb/Ch. LPDDR4)} \\
        {tRCmin} & Min. Act. CMD Interval & {60ns (LPDDR4)} \\ 
        {MAC$_{\text{tREFI}}$} & Max. \# of Act. CMDs in tREFI & {255 (MR4 4x)} \\
        {MAC$_{\text{tREFW}}$} & Max. \# of Act. CMDs in tREFW & {2,095K (MR4 4x)} \\ 
        {RH$_{\text{TH}}$} & Th. \# of Act. for RH-Induced Bit-Flip & {20K \cite{frigo2020trrespass}} \\ 
        {\# of Rows/Bank} & \# of Rows/Bank & {64K (8Gb/Ch. LPDDR4)} \\ 
        {\# of Banks} & Total \# of Banks per Chip & {8 (LPDDR4)} \\ \hlinewd{1.2pt}
        \end{tabular}
        \begin{tablenotes}[flushleft]
            \item[*] Note that the nominal value of tREFI, tREFW, tRFC, and tRCmin can slightly differ for LPDDR, DDR, GDDR, and HBM. The maximum number of activate commands within tREFW (MAC$_{\text{tREFW}}$) can be calculated as follows: $\text{MAC}_\text{tREFW}=\frac{\text{tREFI-tRFC}}{\text{tRCmin}}\times{\text{8K}}$
        \end{tablenotes}
        \end{threeparttable}
    \end{adjustbox}
\end{table}

\section{RowBleed and RowHammer}\label{sec:2}
\begin{figure}[h]
    \centering
    \vspace{-1em}
    \includegraphics[trim=3cm 0 2.5cm 0 width=1\columnwidth]{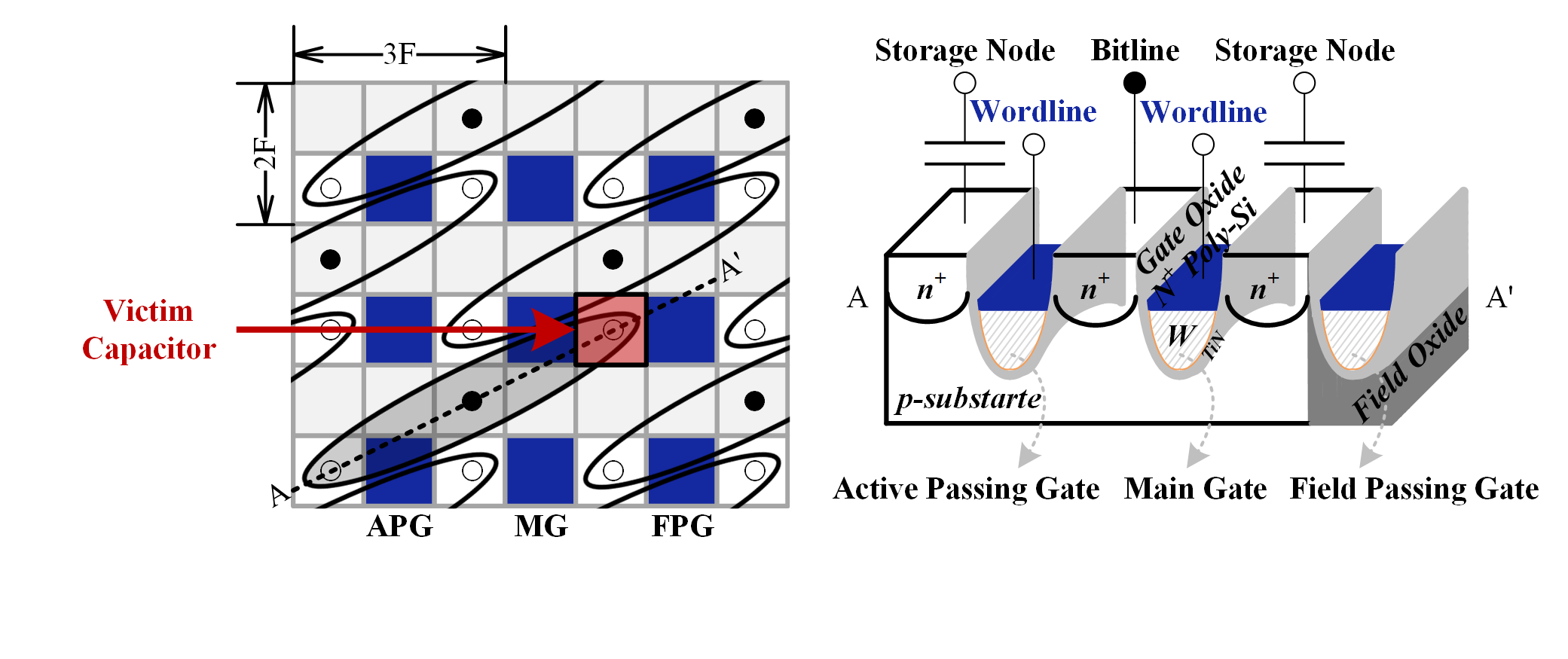}
    \vspace{-2.5em}
    \caption{Physical Cell Layout of Modern DRAM}
    \label{fig:layout}
\end{figure}

Figure \ref{fig:layout} represents the modern DRAM's $6\text{F}^2$ physical cell layout \cite{james2010recent, eetimes}, where F represents half of the bitline pitch. A wordline sharing the bitline with the Main Gate (MG) is called the Active Passing Gate (APG) and a wordline in the field oxide region is called the Field Passing Gate (FPG). This paper assumes the MG as the victim row and demonstrates the physical mechanism of row activation-induced bit-flips from the perspective of the MG. 

\subsection{RowBleed Mechanism}
When either the APG or the FPG is activated for a long period of time, such as in cases where the host system adopts an open-page policy \cite{blackmore2013quantitative}, it acts as an aggressor row that can flip the data of the MG. This phenomenon is referred to as Passing Gate Effect \cite{kim2014architectural}. This paper names this phenomenon as RowBleed to associate with RowHammer.

\begin{figure}[h]
    \centering
    \includegraphics[width=1\columnwidth]{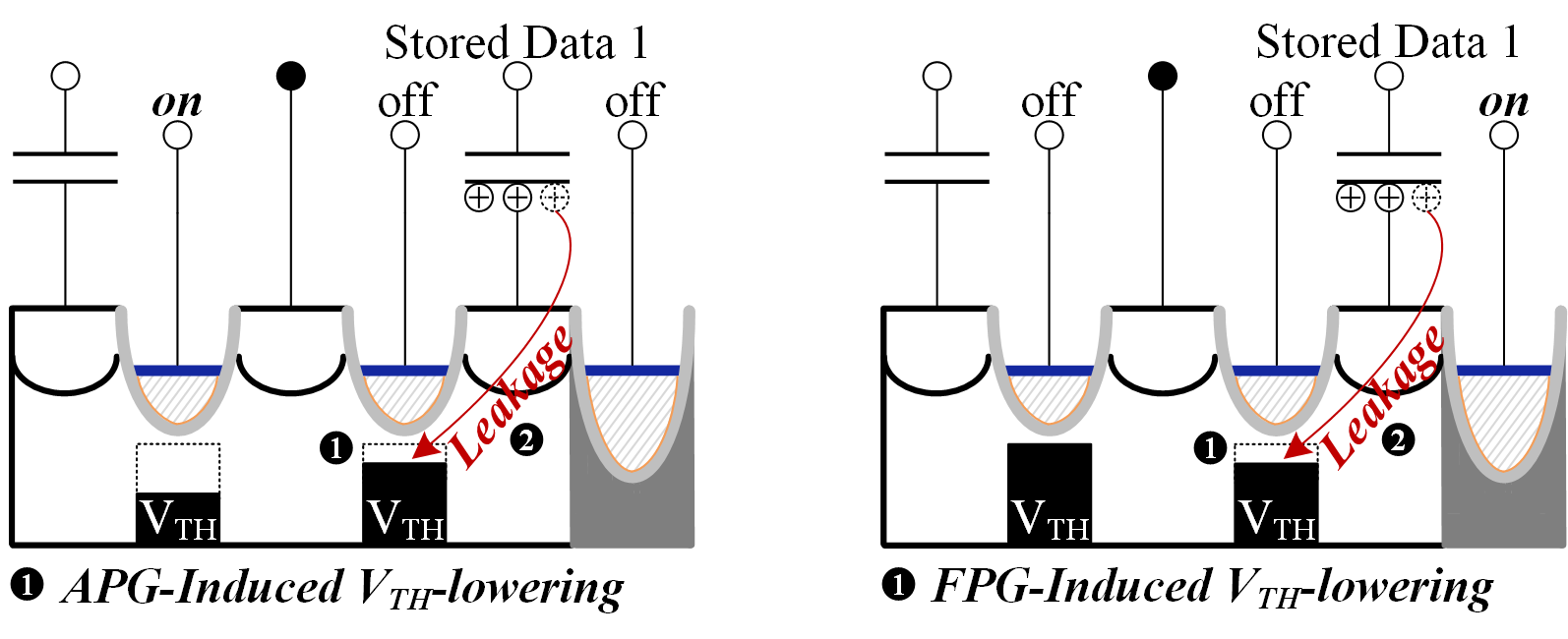}
    \vspace{-2em}
    \caption{RowBleed-Induced Bit-Flip of Stored Data 1}
    \label{fig:pged1}
\end{figure}

Figure \ref{fig:pged1} depicts the RowBleed-induced bit-flip mechanism. When the APG or the FPG is activated, both have high voltage. This high voltage lowers the MG's threshold voltage (V$_{\text{TH}}$), which in turn causes charge \textit{leakage}. Since the transistor's V$_{\text{TH}}$ decreases at high temperature and with cell shrinkage, RowBleed is vulnerable to high temperature and DRAM cell scaling. Note that RowBleed cannot flip stored data 0 due to the direction of the leakage, which is inherent to the nature of the transistor.

\begin{figure}[h]
    \centering
    \includegraphics[trim=1.5cm 0 2.5cm 0 width=1\columnwidth]{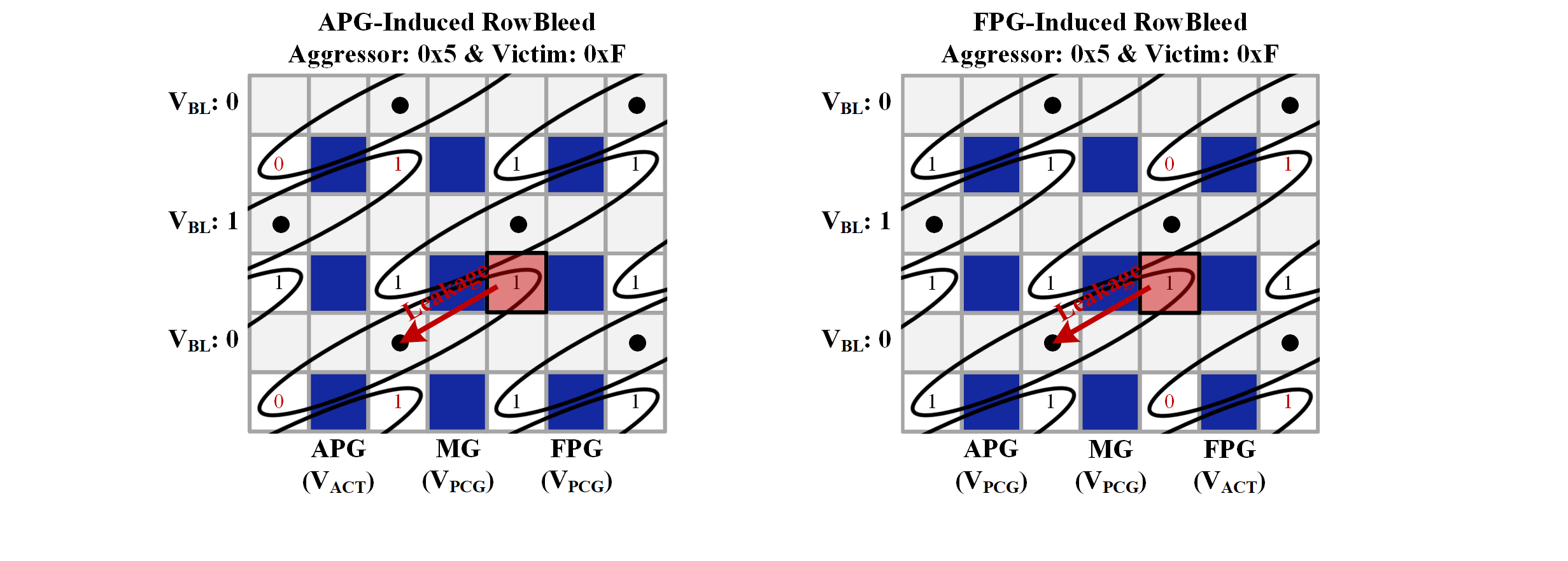}
    \vspace{-2em}
    \caption{RowBleed Accelerating Data Pattern}
    \label{fig:pgepat}
\end{figure}

Figure \ref{fig:pgepat} depicts the RowBleed accelerating data pattern. To accelerate the leakage, a higher voltage difference between the victim capacitor and the bitline can be applied by setting the aggressor row's bitline data to 0x5. Note that V$_{\text{ACT}}$ is the wordline voltage when the wordline is activated, and V$_{\text{PCG}}$ is the wordline voltage when the wordline is precharged.

\begin{figure}[h]
    \centering
    \includegraphics[width=\columnwidth]{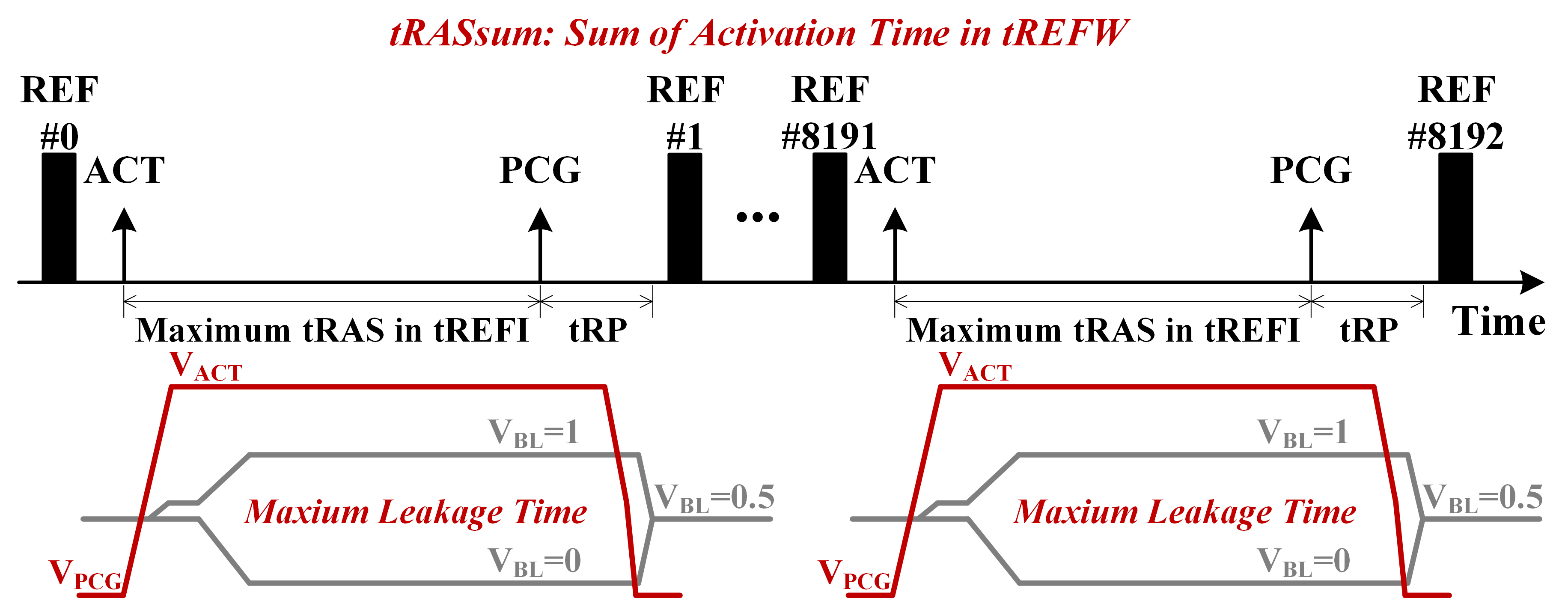}
    \vspace{-2em}
    \caption{RowBleed Accelerating Timing Condition, tRASsum}
    \label{fig:trassum}
\end{figure}

Figure \ref{fig:trassum} depicts the RowBleed accelerating timing condition. To maximize the leakage, the maximum row activation time can be applied. This row activation time is defined as tRAS and it ranges from 42ns to 70,200ns according to memory standard specifications \cite{specification2014mjedec, specification2019jedec, specification2014jedec, specification2020jedec, specification2021gjedec, specification2021hjedec}. Since the aggressor row needs to be in a precharge state while DRAM is refreshed, tRAS cannot be longer than tREFI. Moreover, the victim row can be refreshed in tREFW. Consequently, the sum of tRAS in tREFW needs to be maximized to maximize the leakage. This paper refers to the sum of tRAS in tREFW as \textit{tRASsum}.



RowBleed-induced bit-flips can increase at high temperature, as high temperature facilitates charge leakage.

\subsection{RowHammer Mechanism}
When either the APG or the FPG is frequently activated, it acts as an aggressor row that can flip the data of the MG. This phenomenon is known as RowHammer. In terms of RowHammer, the number of row activations is commonly regarded as a primary factor. This paper, however, demonstrates that row precharge-to-activation time is also a significant factor.

\begin{figure}[h]
    \centering
    \includegraphics[width=1\columnwidth]{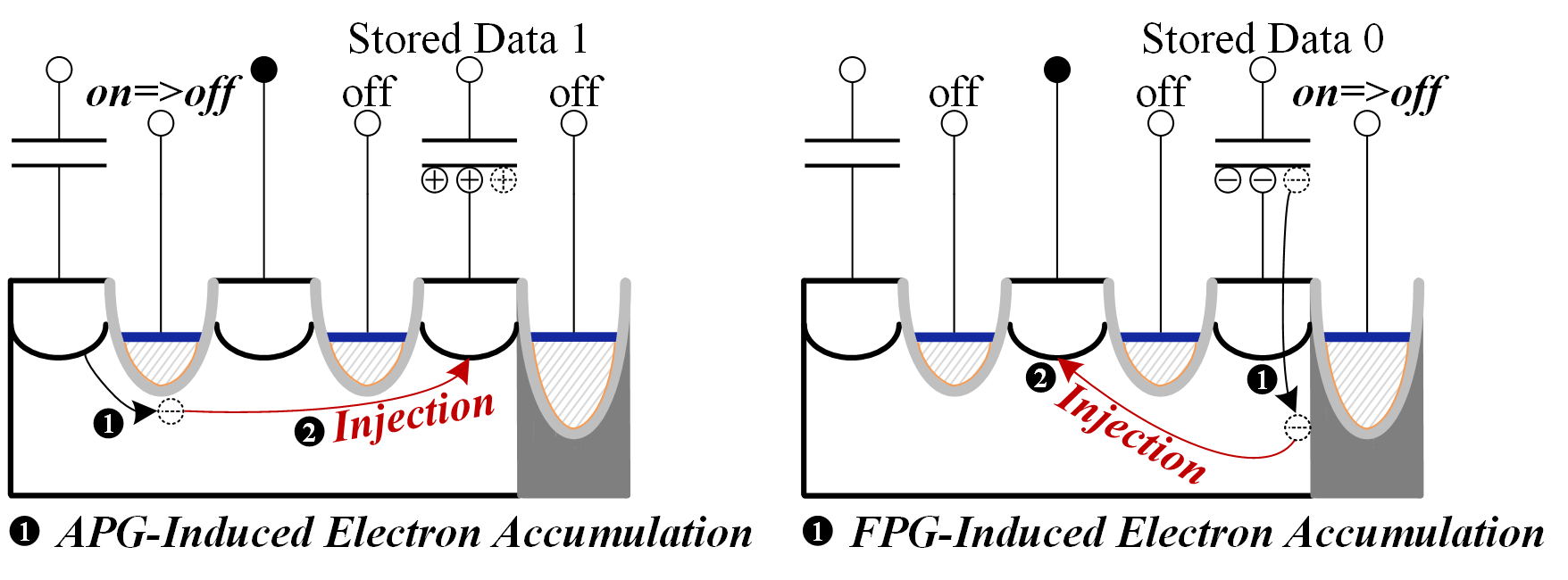}
    \vspace{-2em}
    \caption{RowHammer-Induced Bit-Flip of Stored Data 0 and 1}
    \vspace{-0.5em}
    \label{fig:rh}
\end{figure}

Figure \ref{fig:rh} depicts the RowHammer-induced bit-flip mechanism. When the APG is activated, electrons accumulate around the APG. After the row activation is finished, these accumulated electrons can disperse, and some of the electrons can be \textit{injected} into the MG's capacitor. Repeating this process can cause data 1 to be flipped. When the FPG is activated, electrons accumulate around the FPG. After the row activation is finished, these accumulated electrons can disperse, and some of the electrons can be \textit{injected} into the MG's capacitor. Repeating this process can cause data 0 to be flipped.

There is a phenomenon known as double-sided RowHammer \cite{frigo2020trrespass}. When both the APG and the FPG are frequently activated, they collectively act as aggressor rows that can flip the data of the MG. These gates further facilitate electron injection into the MG by aiding in the transfer of injected electrons.

\begin{figure}[h]
    \centering
    \includegraphics[trim=1.5cm 0 2.5cm 0 width=1\columnwidth]{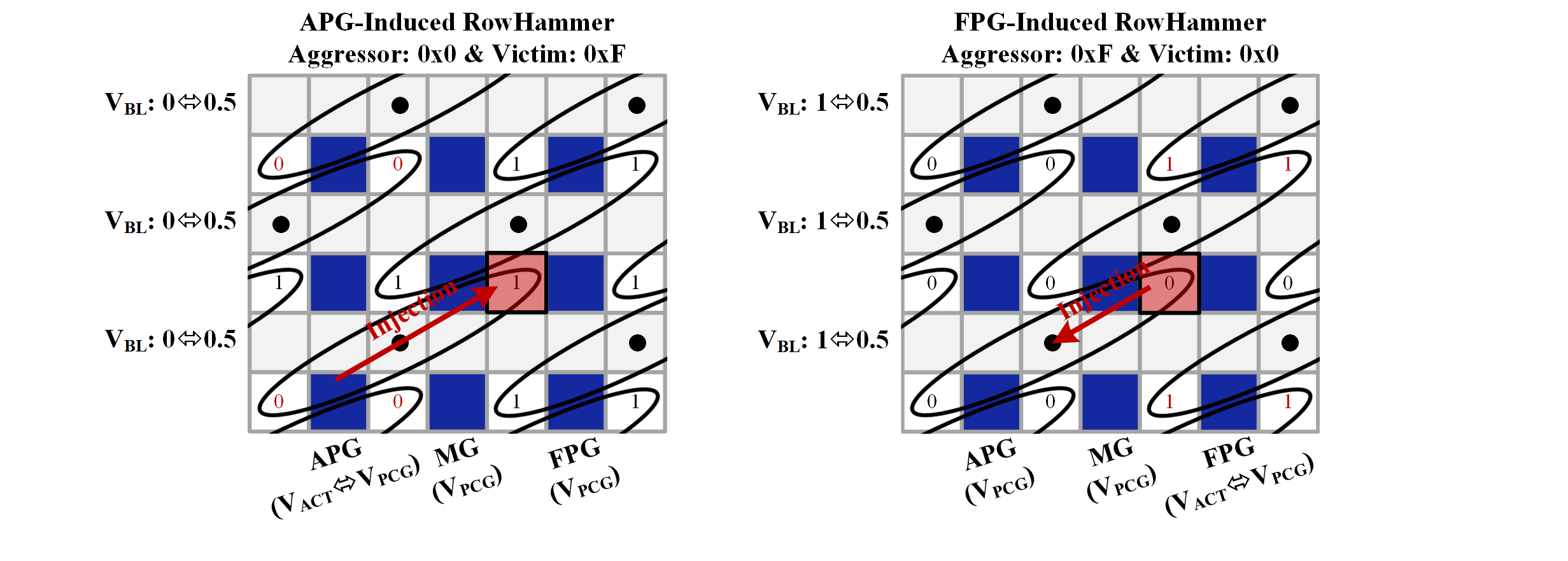}
    \vspace{-1.6em}
    \caption{RowHammer Accelerating Data Pattern}
    \label{fig:rhpat}
\end{figure}

Figure \ref{fig:rhpat} depicts the RowHammer accelerating data pattern. When the MG stores data 1 (or 0), a high voltage difference between the victim capacitor and the bitline is applied to accelerate electron injection from the APG (or FPG) to the MG's capacitor. This is accomplished by setting the APG's (or FPG's) bitline data to 0x0 (or 0xF).

\begin{figure}[h]
    \centering
    \includegraphics[width=\columnwidth]{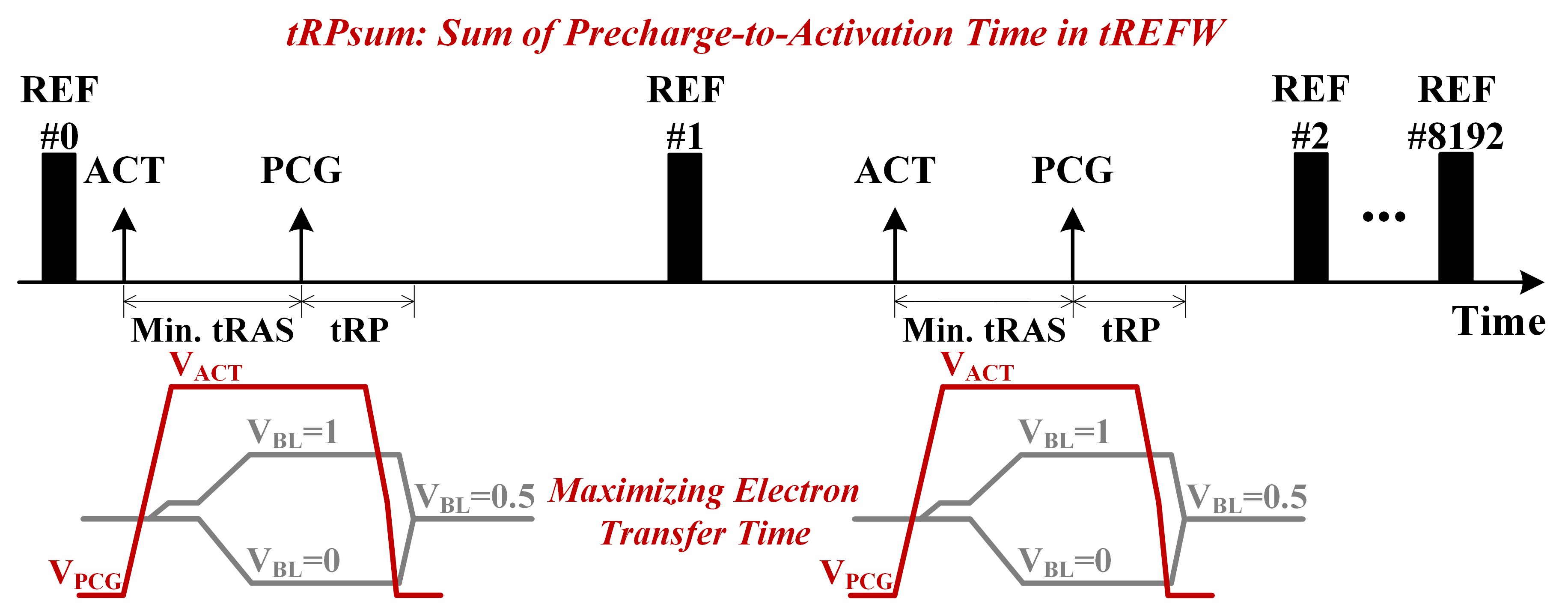}
    \vspace{-2em}
    \caption{RowHammer Accelerating Timing Condition, tRPsum}
    \label{fig:trpsum}
\end{figure}

Figure \ref{fig:trpsum} depicts the RowHammer accelerating timing condition. To maximize the electron accumulation, the minimum tRAS can be applied. However, for electron injection to occur, electrons need to be transferred. This transfer time can be extended by maximizing row precharge-to-activation time. Consequently, the sum of row precharge-to-activation time in tREFW needs to be maximized to maximize electron injection. This paper refers to the sum of row precharge-to-activation time in tREFW as \textit{tRPsum}. Note that row precharge-to-activation time does not stand for the row precharge time defined as tRP in memory standard specifications \cite{specification2014mjedec, specification2019jedec, specification2014jedec, specification2020jedec, specification2021gjedec, specification2021hjedec}.


When the row precharge-to-activation time is short, the number of bit-flips in victim data 0 can exceed that of victim data 1 because victim data 0 is influenced by the FPG, where the physical distance of injection is shorter than in the APG. However, as the row precharge-to-activation time increases, the number of bit-flips in victim data 1 can surpass that of victim data 0 due to the extended injection time. 

RowHammer-induced bit-flips are observed across the entire DRAM operating temperature range \cite{yang2019trap, park2014active}, and the vulnerability is contingent upon the physical characteristics of the cell.


RowHammer-induced bit-flip is not limited to its nearest two rows ($\pm$1), it can affect farther rows ($\pm$2) \cite{walker2021dram}. This paper names this phenomenon as Extended RowHammer. 

\begin{figure}[h]
    \centering
    \includegraphics[trim=1.5cm 0 2.5cm 0 width=1\columnwidth]{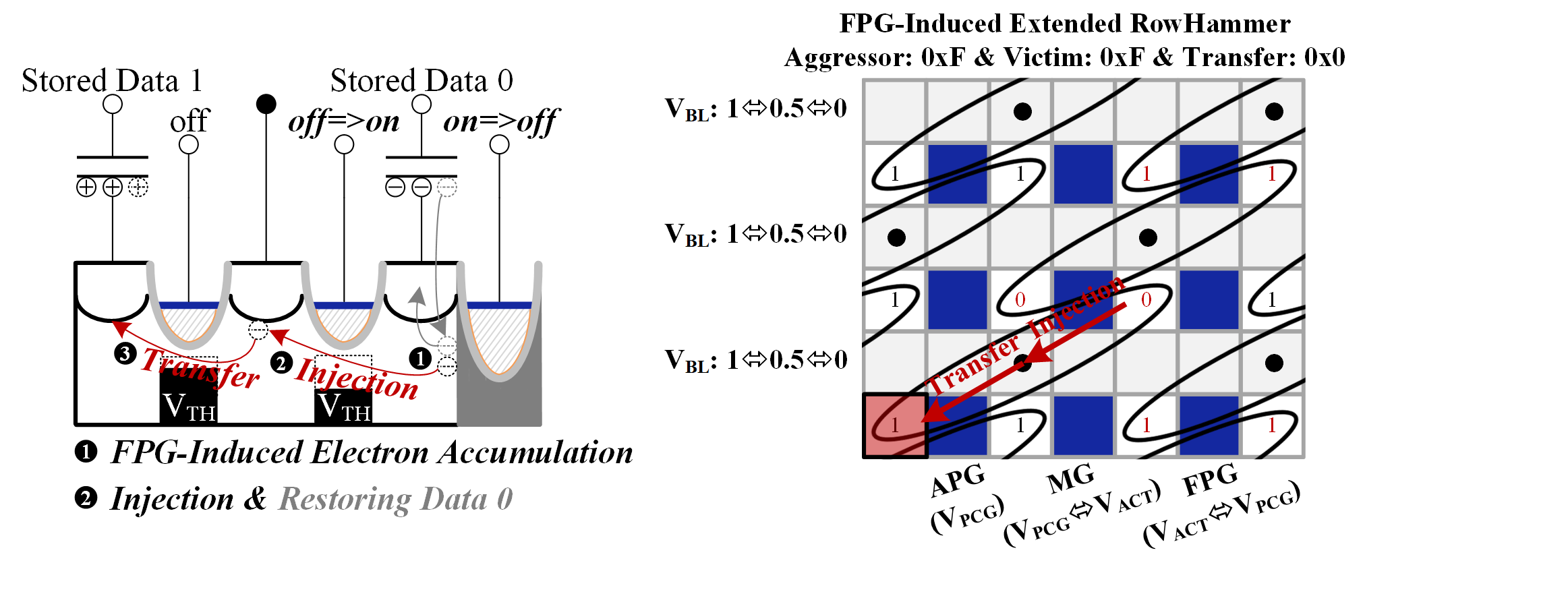}
    \vspace{-1.7em}
    \caption{Extended RowHammer-Induced Bit-Flip of Stored Data 1 \& Extended RowHammer Accelerating Data Pattern}
    \label{fig:erh}
\end{figure}

Figure \ref{fig:erh} depicts the Extended RowHammer bit-flip mechanism. When the FPG is activated, electrons from both the silicon substrate and the MG's capacitor accumulate around the FPG. After the row activation is finished, these accumulated electrons can disperse, and some of the electrons can be \textit{injected} into the silicon substrate. When the MG is activated, the MG's capacitor restores its data 0 and \textit{transfers} the injected electrons to the APG's capacitor. Repeating this process can cause data 1 to be flipped. Note that Extended RowHammer cannot flip stored data 0 since the activated FPG cannot accumulate holes to inject. To accelerate electron injection, a high voltage difference between the MG's capacitor and the bitline is applied by setting the FPG's bitline data to 0xF. To accelerate electron transfer, a high voltage difference between the APG's capacitor and the bitline is applied by setting the MG's bitline data to 0x0.

Extended RowHammer-induced bit-flips can also increase as the row precharge-to-activation time extends, for the same reasons as in RowHammer-induced bit-flips. Extended RowHammer-induced bit-flips can increase in the presence of RowHammer, as RowHammer facilitates the transfer of injected electrons. 

However, due to the physical cell layout (Figure \ref{fig:layout}), the victim row is confined within +/-1 and +/-2 beyond aggressor rows. Furthermore, only the FPG can act as an aggressor since the APG cannot influence the cell of the FPG, where the FPG assumes the role of the MG from that cell's perspective. Additionally, victim data is limited to 1 as the activated FPG cannot accumulate holes for injection.

Table \ref{tab:rbrh} presents a summary of the characteristics of RowBleed and RowHammer. 

\begin{table}[h]
\vspace{-1ex}
\centering
\caption{Summary of RowBleed and RowHammer}
\vspace{-1ex}
\label{tab:rbrh}
    \begin{adjustbox}{width=\columnwidth}
        \begin{threeparttable}
        \begin{tabular}{cccc}
        \hlinewd{1.2pt}
        {}                    & {RowBleed}       & {RowHammer}          & {Extended RowHammer} \\ \hline
        {Mechanism}           & {Charge Leakage} & {Electron Injection} & {Electron Injection} \\
        {Aggressor Row}       & {APG \& FPG}     & {APG \& FPG}         & {FPG} \\
        {Victim Row Data}     & {1}              & {0 \& 1}             & {1} \\
        {Accelerating Temp.}  & {High Temp.}     & {Depends on Cells}   & {Depends on Cells} \\
        {Accelerating Timing} & {tRASsum}        & {tRPsum}             & {tRPsum} \\ \hlinewd{1.2pt}
        \end{tabular}
        \end{threeparttable}
    \end{adjustbox}
\end{table}

\section{DSAC}\label{sec:4}
\subsection{RowBleed Countermeasure}
\noindent\textbf{Time-Weighted Counting.} Since RowBleed can occur when a row is activated for a long period of time, the row activation time (tRAS) which ranges from 42ns to 70,200ns according to memory standard specifications \cite{specification2014mjedec, specification2019jedec, specification2014jedec, specification2020jedec, specification2021gjedec, specification2021hjedec} is a crucial factor.

\begin{figure}[h]
    \centering
    \includegraphics[width=0.8\columnwidth]{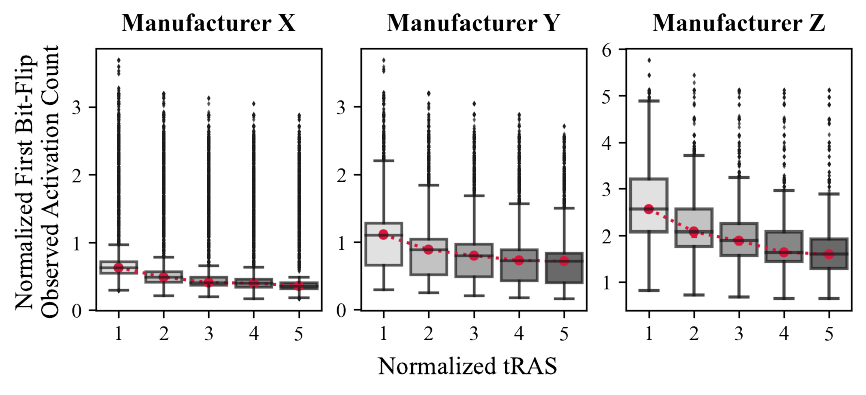}
    \vspace{-1em}
    \caption{Bit-Flip Characteristic of Different tRAS}
    \label{fig:tras}
\end{figure}

Figure \ref{fig:tras} indicates that the relationship between tRAS and bit-flip threshold is nonlinear, as the gradient becomes gradual when tRAS increases. Therefore, this paper proposes the first in-DRAM RowBleed mitigation mechanism named Time-Weighted Counting. This algorithm utilizes a logarithmic function to increase counter weight when tRAS is longer than minimum tRAS (tRASmin) defined by memory standard specifications \cite{specification2014mjedec, specification2019jedec, specification2014jedec, specification2020jedec, specification2021gjedec, specification2021hjedec}, while the growth slows down as tRAS increases. The function can be expressed as follows:
\begin{equation}
\centering
W_C=\mathrm{\alpha\times\log_2\frac{tRAS}{tRASmin}}
\label{eq2}
\end{equation}, where $W_C$ is a counter weight for each row, representing the extent of RowBleed-induced leakage, and $\alpha$ is a coefficient that can be fine-tuned. If $\alpha$ is greater than 0 and tRAS equals tRASmin, $W_C$ becomes 0. However, if $\alpha$ is greater than 0 and tRAS equals $2\times\text{tRASmin}$, $W_C$ becomes $\alpha\times1$. This weight is added to the corresponding row's count value in a count table. 

Consequently, if multiple rows are activated the same number of times, but one row is activated for a longer duration than the others, its count value will surpass those of the others. As a result, DRAM can prioritize refreshing that row first through TRR.

\noindent\textbf{Dynamic Body-Bias.} Dynamic Body-Bias \cite{tschanz2003dynamic} can be leveraged for RowBleed mitigation by dynamically adjusting the body voltage of a victim row based on the state of its adjacent aggressor row. The core concept is that lowering the body voltage of a victim row can compensate for the threshold voltage lowering caused by its adjacent aggressor row.

Lowering V$_{\text{ACT}}$ can be considered to mitigate RowBleed since high V$_{\text{ACT}}$ can accelerate V$_{\text{TH}}$-lowering. However, lowering V$_{\text{ACT}}$ has a negative impact on data-write time due to the high capacitance of DRAM cell capacitor. Lowering V$_{\text{PCG}}$ can also be considered to mitigate RowBleed since high V$_{\text{PCG}}$ can accelerate V$_{\text{TH}}$-lowering. However, lowering V$_{\text{PCG}}$ has a negative impact on RowHammer due to a higher voltage difference between V$_{\text{PCG}}$ and V$_{\text{ACT}}$ can accelerate RowHammer disturbance on electrons.


As an alternative, the body voltage of the MG (V$_{\text{BODY}}$) can be lowered when its neighboring row is activated. Since lowering V$_{\text{BODY}}$ can increase the MG's V$_{\text{TH}}$, it helps compensate for the V$_{\text{TH}}$-lowering impact by the aggressor row.

Dynamic Body-Bias \cite{tschanz2003dynamic} can be implemented in the global wordline driver to control the source voltage of the local wordline NMOS in the sub-wordline driver. However, this requires the generation of additional voltage levels, which increases test time—an important consideration for mass production. In contrast, Time-Weighted Counting can be implemented more simply, as it leverages the existing in-DRAM RowHammer counter. Therefore, this paper adopts Time-Weighted Counting.

\noindent\textbf{CAS-Only DRAM.} Eliminating explicit active and precharge commands in DRAM can reduce the risk of row activation-induced bit-flips. Instead, CAS commands, such as write and read with auto-precharge, can be employed for row activation and precharge. While CAS-only DRAM increases data write and read latency, it can maintain data bandwidth by increasing the number of DQ pins and banks. To further improve data write and read latency, smaller cell arrays are needed in DRAM banks. However, this would require significant modifications to memory standard specifications. Therefore, this paper leaves this topic for future research.

\subsection{RowHammer Countermeasure}
Since RowHammer can occur when a row is frequently activated, row activation count is a major factor. Identifying the most frequently appearing elements is an active research area \cite{anderson2017high, chen2002multi, cormode2008finding, cormode2010methods, cormode2005s, cormode2005improved, giannella2003mining, lim2017time, liu2013data, manerikar2009frequent, metwally2006integrated, qureshi2007adaptive, zadnik2011evolution, krishnamurthy2003sketch, hua2008brick, tong2015high, callegari2011detecting, schweller2004reverse}. This paper focuses on counter-based algorithms due to their low space complexity. 

\noindent\textbf{Approximate Counting.} Among counter-based algorithms \cite{misra1982finding, manku2002approximate, metwally2006integrated, cormode2008finding, cormode2010methods, manerikar2009frequent, anderson2017high}, Space Saving algorithm is widely considered the most area-efficient detection algorithm \cite{cormode2008finding, cormode2010methods, manerikar2009frequent, anderson2017high}. 

Space Saving algorithm updates its count table for every incoming row. The key idea of the algorithm is to keep track of a replaced row's count so that it can approximate the replaced row's count value when it returns to a count table. For example, if a minimum count row(\textit{y}) is replaced by a new row(\textit{x}), the count value for \textit{x} becomes count(\textit{y})$+$1, rather than discarding count(\textit{y}). By leveraging this Approximate Counting, Space Saving algorithm can be area-efficient.

However, state-of-the-art counter-based algorithms have a drawback in the context of RowHammer detection. Specifically, these detection algorithms are vulnerable to decoy-rows due to DRAM's limitation on the number of counters for the TRR algorithm. Decoy-rows refer to rows whose number of accesses does not exceed the number of RowHammer accesses within an observation period, and should not be inserted into the detection algorithm's count table.

\begin{figure}[h]
    \centering
    \includegraphics[width=1\columnwidth]{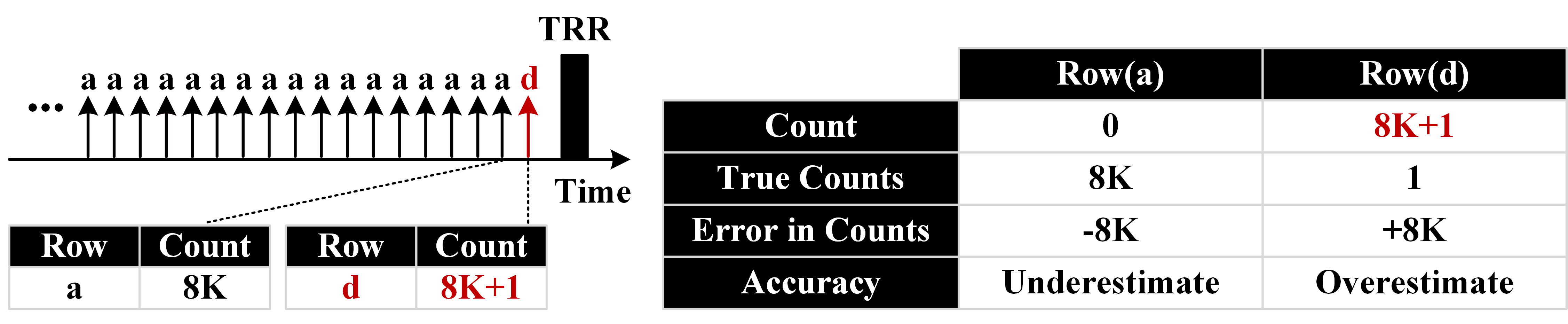}
    \vspace{-2em}
    \caption{Problem of State-of-the-Art Counter-Based Algorithms}
    \label{fig:problem}
\end{figure}

Figure \ref{fig:problem} simplifies the drawback of Space Saving algorithm. There is only one counter available, and TRR is performed on the black bar labeled as TRR. There are two incoming rows: the aggressor-row(a) and the decoy-row(d). Since the Space Saving algorithm updates its count table for every incoming row, the decoy-row(d) takes all the count value of the aggressor-row(a). As a result, the aggressor-row(a) is significantly underestimated, while the decoy-row(d) is overestimated. Consequently, the victim rows of the aggressor-row(a) cannot be TRRed, which can lead to RowHammer-induced bit-flips.

As shown above, the detection performance of Space Saving algorithm strongly depends on the number of counters. The error in counts (\textit{Ce}) can be represented as follows:
\begin{equation}
\centering
Ce<\lfloor\frac{n}{c}\rfloor
\label{eq3}
\end{equation}, where \textit{n} is the number of row activations and \textit{c} is the number of counters. Note that \textit{Ce} can be maximized when the number of rows is equal to the number of counters$+$1.

\begin{figure}[h]
    \centering
    \includegraphics[width=\columnwidth]{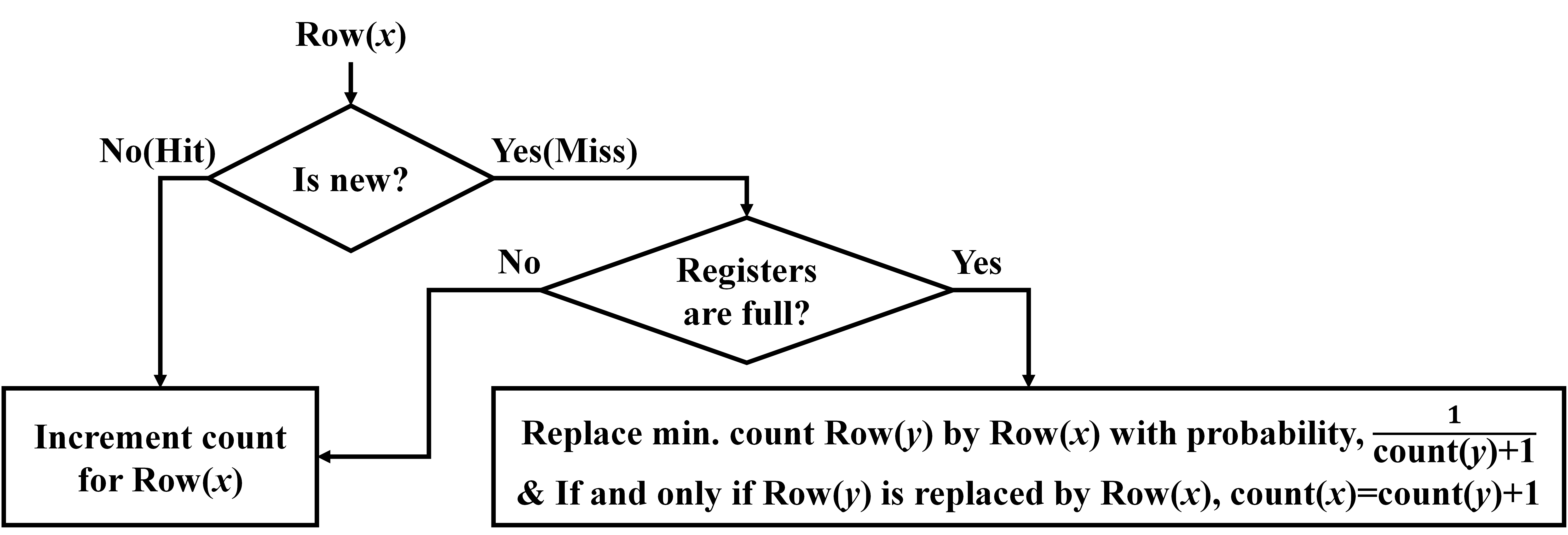}
    \vspace{-2em}
    \caption{Flowchart of DSAC}
     \label{fig:flow}
\end{figure}

\noindent\textbf{Stochastic Replacement.} Based on the above analysis, this paper focuses on minimizing \textit{Ce} by filtering out decoy-rows. Since the number of decoy-row accesses cannot exceed the number of RowHammer accesses within an observation period, an algorithm is required to filter out rows whose number of counts is lower than the number of RowHammer accesses. Therefore, this paper proposes DSAC, which filters out decoy-rows by adopting Stochastic Replacement. Figure \ref{fig:flow} shows the flowchart of DSAC.

\begin{itemize}[leftmargin=*,align=left, noitemsep, topsep=0pt, parsep=0pt]
\item Hit: Corresponding row's count value is incremented.
\item Miss: A count table is checked to see if it is full.
\item Insertion: If count table is not full, a new row is inserted into a count table, and its count value is incremented.
\item Replacement: If a count table is full, a row with the minimum count value can be replaced by a new row. The probability of selecting the row to replace is determined by \begin{equation}\text{P}(r)=\frac{1}{\text{min. cnt} + 1}\label{eq5}\end{equation}, where min. cnt is the minimum count value in a count table, and the minimum count value is incremented by 1 if a replacement occurs.
\end{itemize}

The key idea of DSAC is to use Stochastic Replacement to replace a new row with a minimum count row in a count table. Specifically, a new row(\textit{x}) can replace a minimum count row(\textit{y}) with a replacement probability, $\text{P}(r) = \frac{1}{\text{count}(y)+1}$. Therefore, row(\textit{x}) can be inserted into a count table if it comes in more than count(\textit{y})+1 on average. Note that DSAC leverages Approximate Counting for area-efficiency by keeping track of the count value of replaced rows.

\begin{figure}[h]
    \centering
    \includegraphics[width=\columnwidth]{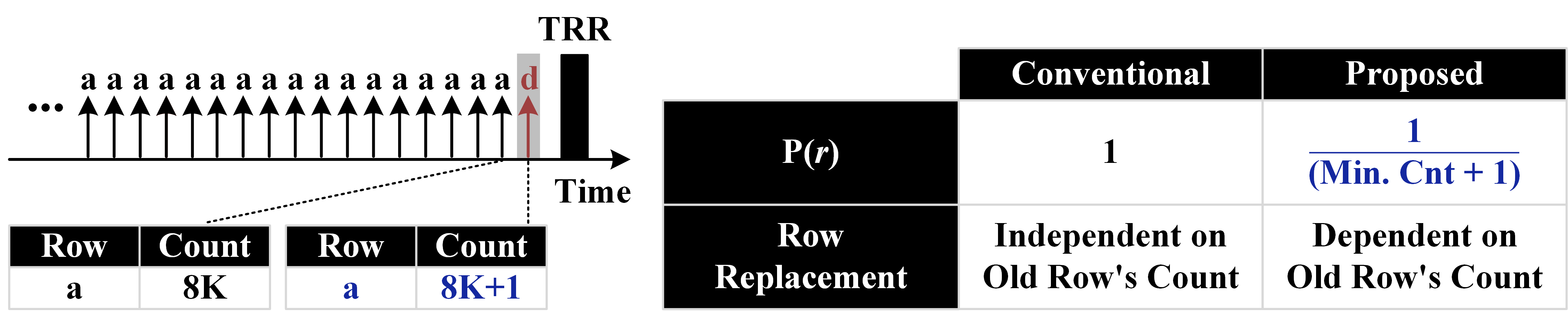}
    \vspace{-2em}
    \caption{High-Level Overview of DSAC}
     \label{fig:high_level}
\end{figure}

Figure \ref{fig:high_level} illustrates how DSAC can solve the problem presented in Figure \ref{fig:problem}. Since $\text{P}(r)=\frac{1}{\text{8K}+1}$ is a low probability, DSAC can keep the aggressor-row(a) in a count table and TRR the neighboring victim rows. As a result, DSAC can minimize \textit{Ce} as follows:
\begin{equation}
\centering
Ce\leq\lfloor\frac{\text{min. cnt}}{c}\rfloor
\label{eq6}
\end{equation}, where min. cnt is the minimum count value in a count table and \textit{c} is the number of counters. A rigorous mathematical analysis supporting this claim is provided in Section \ref{sec:5}. 

\begin{figure}[h]
    \centering
    \includegraphics[width=\columnwidth]{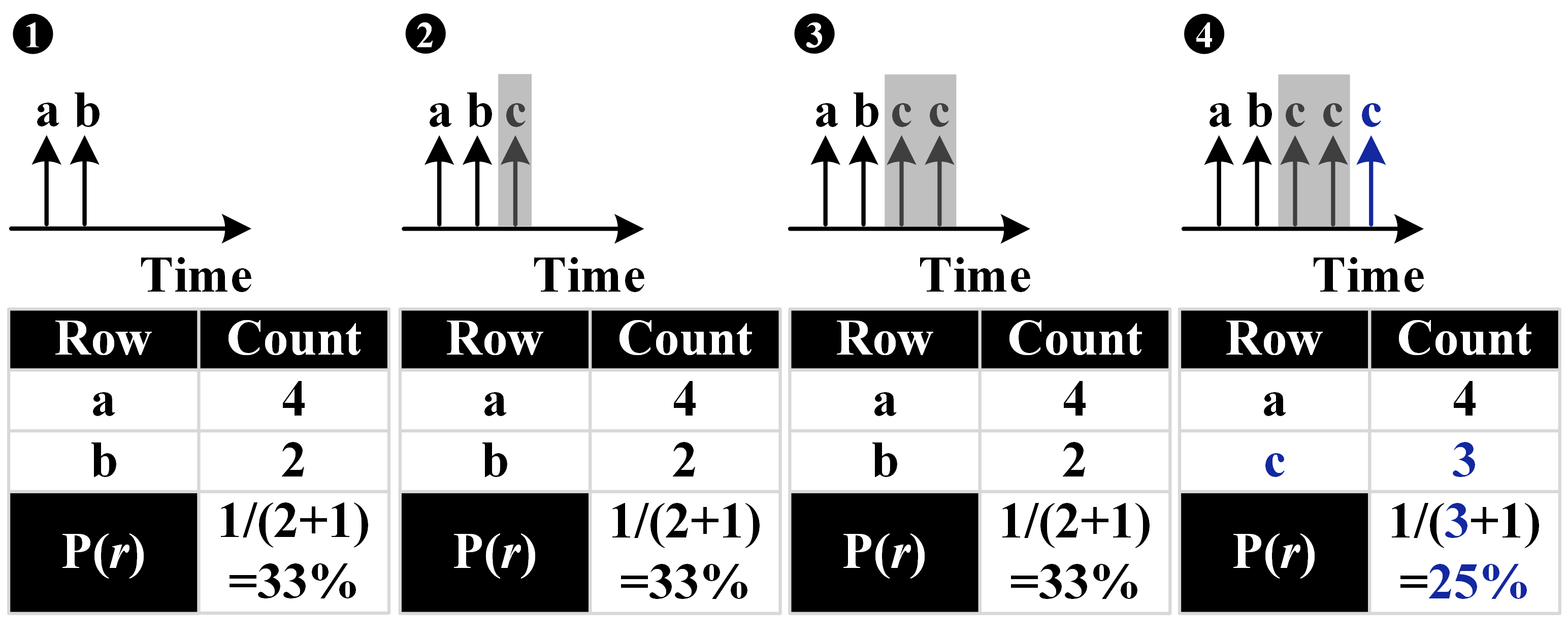}
    \vspace{-2em}
    \caption{Operation Example of 2 Count Table DSAC}
     \label{fig:reg_2}
\end{figure}

Figure \ref{fig:reg_2} illustrates how DSAC operates when the number of counters is equal to 2. {{\small \bcirc{1}}} A count table is full with row(a) and row(b), and the minimum count value equal is 2. This sets P(\textit{r}) equal to 1/(2$+$1), which is 33\%. {{\small \bcirc{2}}} New row(c) comes in, yet it does not replace old row(b). Hence, P(\textit{r}) remains the same. {{\small \bcirc{3}}} Row(c) comes in again, but still does not replace old row(b), so P(\textit{r}) remains the same. {{\small \bcirc{4}}} Row(c) replaces old row(b), so the minimum count value is set to 3, and P(\textit{r}) is set to 25\%. Note that P(\textit{r}) can be reset to 1 after TRR.

In summary, this paper proposes DSAC for filtering out decoy-rows with area-efficiency. The pseudocode of DSAC can be found in Algorithm \ref{alg:alg}.

\newcommand\mycommfont[1]{\scriptsize\textcolor{black}{#1}}
\SetCommentSty{mycommfont}

\begin{algorithm}[h]
\scriptsize
\SetAlCapNameFnt{\scriptsize}
\SetAlCapFnt{\scriptsize}
\SetAlgorithmName{Algorithm}
\DontPrintSemicolon

    \tcp{Hit}
    \If{{incoming$\_$ROW == count$\_$table[i][`ROW']}}
    {count$\_$table[i][`CNT'] ++\\\algorithmicreturn
    }
    \tcp{Miss}
    \Else{
        \tcp{Insertion}
        \If{count$\_$table[i][`ROW'] == None}
        {count$\_$table[i][`ROW'] = Incoming$\_$ROW\\count$\_$table[i][`CNT'] ++\\\algorithmicreturn}
        \tcp{Replacement}
        \Else{
        i\quad\quad\quad\space= argmin$_\text{j}$\text{(count$\_$table[j][`CNT'])}\\
        min$\_$cnt = min(count$\_$table[j][`CNT'])\\
        \If{RANDOM[0,1) $\leq$ 1 / (min$\_$cnt + 1)}
        {count$\_$table[i][`ROW'] = Incoming$\_$ROW\\count$\_$table[i][`CNT'] ++\\\algorithmicreturn}
        \Else{\algorithmicreturn}
            }
            
        }
\vspace{-0.25em}
\caption{Pseudocode of DSAC}
\label{alg:alg}
\end{algorithm}

\noindent\textbf{TRR Threshold.} TRR can be performed on a refresh command that can be issued after DSAC reaches its TRR threshold (TRR$_{\text{TH}}$). Since the number of row activations within tREFI can vary, DSAC introduces adaptive TRR$_{\text{TH}}$ which can change TRR$_{\text{TH}}$ depending on the sum of counts in DSAC's count table. Flag for TRR$_{\text{TH}}$ is triggered according to Inequality \ref{eq7}. 

\begin{equation}
\centering
\text{Sum of Counts in Count Table}\geq\frac{\text{RH}_\text{TH}}{2}-\text{MAC}_\text{tREFI}
\label{eq7}
\end{equation}, where $\frac{\text{RH}_\text{TH}}{2}$ is employed to mitigate double-sided hammer, a scenario where a victim row is positioned between two aggressor rows \cite{frigo2020trrespass}. In such cases, the victim row requires to be TRRed before one of the aggressor rows reaches a count value equal to $\frac{\text{RH}_\text{TH}}{2}$. Note that different TRR$_{\text{TH}}$ can also be adopted if necessary.

\subsection{Security Analysis}\label{sec:5}
Since DSAC leverages probability, the security analysis of DSAC is based on probability theory. In order to guarantee DSAC's security, this paper demonstrates the worst attack pattern and proves its security against the worst attack pattern.

\noindent\textbf{Theorem.} \textit{DSAC can guarantee its security against double-sided uniform weight pattern, where a victim row is positioned in between two aggressor rows \cite{frigo2020trrespass} and all the incoming rows have uniformly distributed weights.}

\noindent\textbf{Lemma 1.} \textit{Double-sided uniform weight pattern is the worst pattern to DSAC.} 

\noindent\textbf{Proof.} Assume non-uniform weight pattern, where independent and identically distributed \textit{k} rows access with random weights, and P(\textit{n}) is the probability of an event, where the corresponding event is when the number of activations (nA) of an arbitrary row exceeds $\frac{\text{RH}_\text{TH}}{2}$. Then, P(\textit{n}) can be expressed as follows: $\text{P}(n)=Cx_1\times\text{P}(\text{nA}>\frac{\text{RH}_\text{TH}}{2}|o_1)+Cx_2\times\text{P}(\text{nA}>\frac{\text{RH}_\text{TH}}{2}|o_2)+\cdot\cdot\cdot+Cx_k\times\text{P}(\text{nA}>\frac{\text{RH}_\text{TH}}{2}|o_k)$, where \textit{o}$_k$ represents an access proportion of row \textit{k} (the number of activations of row \textit{k} / the total number of activations), \textit{Cx}$_k$ represent the number of rows that have a proportion of \textit{o}$_k$, and $\text{P}(\text{nA}>\frac{\text{RH}_\text{TH}}{2}|o_k)$ represents the probability of an event whose number of activations of row with \textit{o}$_k$ exceed $\frac{\text{RH}_\text{TH}}{2}$.

Assume that $Cx_1\times\text{P}(\text{nA}>\frac{\text{RH}_\text{TH}}{2}|o_1)\geq Cx_2\times\text{P}(\text{nA}>\frac{\text{RH}_\text{TH}}{2}|o_2)\geq\cdot\cdot\cdot\geq Cx_k\times\text{P}(\text{nA}>\frac{\text{RH}_\text{TH}}{2}|o_k)$, then $(\frac{C_1}{C_2}\times\text{P}(\text{nA}>\frac{\text{RH}_\text{TH}}{2}|o_1)\geq\text{P}(\text{nA}>\frac{\text{RH}_\text{TH}}{2}|o_2)$, $(\frac{C_1}{C_3}\times\text{P}(\text{nA}>\frac{\text{RH}_\text{TH}}{2}|o_1)\geq\text{P}(\text{nA}>\frac{\text{RH}_\text{TH}}{2}|o_3)$, and $(\frac{C_1}{C_k}\times\text{P}(\text{nA}>\frac{\text{RH}_\text{TH}}{2}|o_1)\geq\text{P}(\text{nA}>\frac{\text{RH}_\text{TH}}{2}|o_k)$. Then, P(\textit{n}) can be expressed as follows: $\text{P}(n)\leq Cx_1\times\text{P}(\text{nA}>\frac{\text{RH}_\text{TH}}{2}|o_1)+Cx_2\times\frac{C_1}{C_2}\times\text{P}(\text{nA}>\frac{\text{RH}_\text{TH}}{2}|o_1)+\cdot\cdot\cdot+Cx_k\times\frac{C_1}{C_k}\times\text{P}(\text{nA}>\frac{\text{RH}_\text{TH}}{2}|o_1)$.

In the case of uniform weight pattern, all the rows have an equal number of activations. Thus, $o_1=o_2=\cdot\cdot\cdot=o_k=\frac{1}{C_1}$. Since $o_1\times Cx_1+o_2\times Cx_2+\cdot\cdot\cdot+o_k\times Cx_k=1$, $\frac{Cx_1}{C1}+\frac{Cx_2}{C_2}+\cdot\cdot\cdot+\frac{Cx_k}{C_k}=1$. Then, P(\textit{n}) can be expressed as follows: $\text{P}(n)\leq [Cx_1+Cx_2\times\frac{C_1}{C_2}+\cdot\cdot\cdot+Cx_k\times\frac{C_1}{C_k}\times\text{P}(\text{nA}>\frac{\text{RH}_\text{TH}}{2}|o_1)=C_1\times\text{P}(\text{nA}>\frac{\text{RH}_\text{TH}}{2}|o_1)$.

In conclusion, P(\textit{n}) can be expressed as follows:
\begin{equation}
\centering
\text{P}(n) \leq C_1\times\text{P}(\text{nA}>\frac{\text{RH}_\text{TH}}{2}|o_1)
\label{eqpk4}
\end{equation}

Inequality \ref{eqpk4} proves that uniform weight pattern is the worst pattern.

\noindent\textbf{Lemma 2.} \textit{Double-sided uniform weight pattern can minimize P(\textit{r}). However, DSAC can detect one of the double-sided aggressor rows.} 

\noindent\textbf{Proof.} If DSAC consecutively filters out one of the double-sided aggressor rows $\frac{\text{RH}_\text{TH}}{2}$ times, then RowHammer-induced bit-flip can occur. Double-sided uniform weight pattern can minimize P(\textit{r}) which can lead to consecutive filtering.

For example, if the number of aggressor rows is greater than the number of counters, then some of the aggressor rows can be filtered out until they access more than a minimum count row in a count table on average. In this case, P(\textit{r}) can be minimized if all the incoming rows have equal weights. If DSAC consecutively filters out one of the aggressor rows $\frac{\text{RH}_\text{TH}}{2}$ times due to the low P(\textit{r}), then RowHammer-induced bit-flip can occur.

However, the probability of $\frac{\text{RH}_\text{TH}}{2}$ consecutive filtering is extremely low due to adaptive TRR$_{\text{TH}}$ in Inequality \ref{eq7}. Since TRR is performed for every refresh command when the sum of counts in a count table is more than $\frac{\text{RH}_\text{TH}}{2}-\text{MAC}_\text{tREFI}$, the minimum count value in a count table cannot exceed $\frac{\text{RH}_\text{TH}/2-\text{MAC}_\text{tREFI}}{\text{number of counters}}$. This sets an upper bound of minimum count value as follows: $m\leq(\frac{\text{RH}_\text{TH}}{2}\text{MAC}_\text{tREFI})/c$, where \textit{m} is the minimum count value in a count table and \textit{c} is the number of counters. Note that the count value is divided by the number of counters since all the counters have uniform weights until TRR is performed or replacement occurs. This sets an lower bound of P(\textit{r}) as follows: 
\begin{equation}
\centering
\text{P}(r)=\frac{1}{m+1}\geq\frac{1}{(\frac{\text{RH}_\text{TH}}{2}-\text{MAC}_\text{tREFI})/c+1}
\label{eq9}
\end{equation}

Using Inequality \ref{eq9}, the probability of $\frac{\text{RH}_\text{TH}}{2}$ consecutive filtering can be expressed as follows:
\begin{equation}
\centering
\begin{aligned}
\text{P}(\textit{f})=&(1-\text{P}(\textit{r}))^{\frac{\text{RH}_\text{TH}}{2}}\\
                    =&(1-\frac{1}{(\frac{\text{RH}_\text{TH}}{2}-\text{MAC}_\text{tREFI})/c+1})^{\frac{\text{RH}_\text{TH}}{2}}
\label{eq10}
\end{aligned}
\end{equation}

Equality \ref{eq10} represents the case of one row filtering where the number of rows is equal to the number of counters$+$1. In order to consider the case of multiple rows filtering, Equality \ref{eq10} can be generalized as follows:
\begin{equation}
\centering
\begin{aligned}
\text{P}(f)=&\sum_{i=1}^{k}Cx_k\times(1-o_i\times\text{P}(r))^{\frac{\text{RH}_\text{TH}}{2}}\\
        \leq&(1-\text{P}(r))^{\frac{\text{RH}_\text{TH}}{2}}
\label{eq11}
\end{aligned}
\end{equation}

Note that Equality \ref{eq11} is bounded by Equality \ref{eq10} for the same reason as explained in Inequality \ref{eqpk4}. Therefore, this paper analyzes the security of DSAC using Equality \ref{eq10}.

\begin{figure}[h]
    \centering
    \includegraphics[width=0.8\columnwidth]{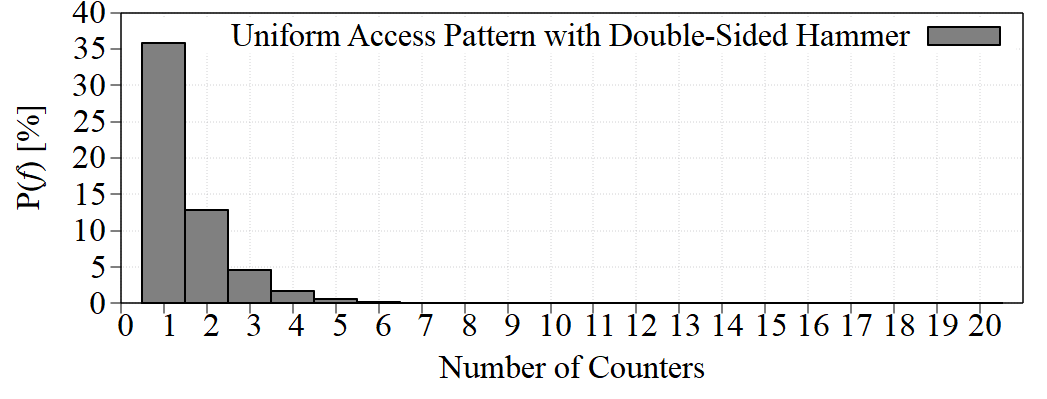}
    \vspace{-1em}
    \caption{Probability of $\frac{\text{RH}_\text{TH}}{2}$ Consecutive Filtering}
    \label{fig:pf}
\end{figure}

Figure \ref{fig:pf} shows that the probability of $\frac{\text{RH}_\text{TH}}{2}$ consecutive filtering is extremely low. Therefore, DSAC can detect one of the double-sided aggressor rows before its count reaches $\frac{\text{RH}_\text{TH}}{2}$. 

\noindent\textbf{Lemma 3.} \textit{In order to make P(\textit{f}) equal to 0, tremendous number of counters is required. However, DSAC can achieve a near-complete detection.}

\noindent\textbf{Proof.} According to Table \ref{tab:base}, 9744 counters are required to make P(\textit{f}) equal to 0. If the number of counters is the same with the state-of-the-art counter-based algorithm named Graphene \cite{park2020graphene} which requires 418 counters ($\text{the required number of counters}=\frac{\text{MAC}_\text{tREFWe}}{(\text{RH}_\text{TH}/4)+1}-1$), then P(\textit{f}) becomes $3.850^{-183}$. This value is approximately 0, yet modern DRAM disallows this many counters due to the area limitation. Therefore, this paper analyzes the security of DSAC with 20 counters as modern DRAM can have for each bank.

If the number of counters is 20, then P(\textit{f}) becomes $1.245^{-9}$. This value appears quite low, yet there is a possibility of failure where filtered rows are never inserted into a count table. However, DRAM has a product lifetime. For example, if DRAM's lifetime is up to 7 years \cite{micronplp}, then P(\textit{f}) does not need to be 0 in perpetuity. To summarize, there is a trade-off between P(\textit{f}) and the number of counters, and they can be determined by the required product lifetime for each application. Therefore, this paper computes the reliability function to show a stochastic product lifetime for P(\textit{f}) to become 1.

This can be done by calculating the complementary cumulative distribution function (CCDF) of the exponential distribution. Equality \ref{eq10} can be interpreted as a geometric distribution since the CCDF of the geometric distribution is as follows: $\text{P}(X>k)=(1-p)^{k}$, where \textit{p} is the probability of success and \textit{k} is the number of trials. While the geometric distribution is in discrete time, the exponential distribution is in continuous time. Hence, if $p=\lambda\tau$, where $\lambda$ is a constant rate parameter equal to $\frac{1}{\text{Mean Time Between Failures}}$ and $\tau$ is a sufficiently small time step, then the geometric distribution approaches the exponential distribution as follows: $\text{P}(X>\frac{x}{\tau})=\lim_{\tau \to 0} (1-\lambda\tau)^{\frac{x}{\tau}}=\lim_{\tau \to 0} [(1-\lambda\tau)^{\frac{1}{\tau}}]^{x}=e^{-\lambda x}$.

Therefore, the reliability function for the exponential distribution can be expressed as follows:
\begin{equation}
\centering
\text{R}(t)=e^{-\lambda t}
\label{eq14}
\end{equation}, where \textit{t} is the lifetime warranty, and $\lambda$ is equal to P(\textit{f}). 

\begin{figure}[h]
    \centering
    \includegraphics[width=0.9\columnwidth]{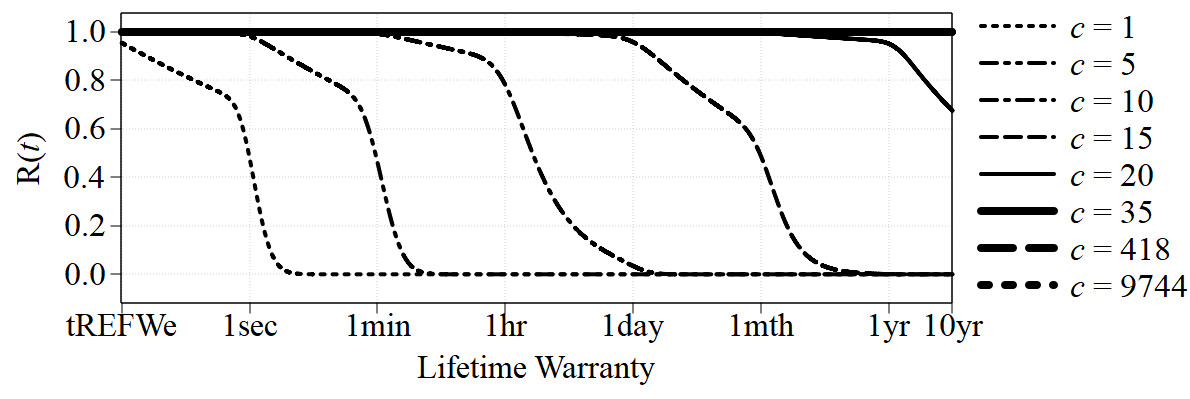}
    \vspace{-1em}
    \caption{Number of Counters Impact on Product Lifetime}
    \label{fig:rt}
\end{figure}

\begin{figure*}[t]
    \centering
    \includegraphics[width=1\textwidth]{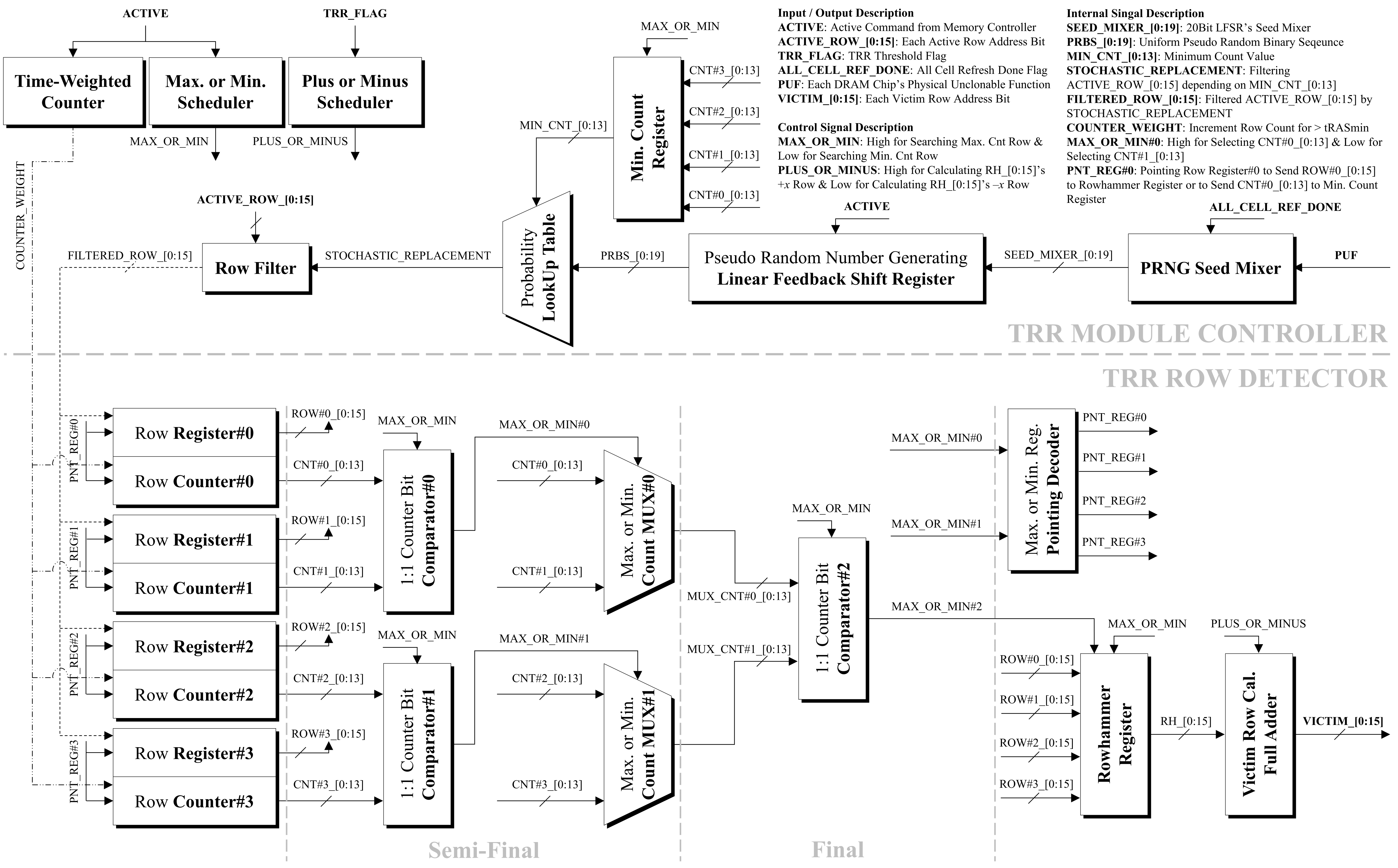}
    \vspace{-1.5em}
    \caption{Architecture of 4 Count Table DSAC with Time-Weighted Counter}
    \vspace{-1.5em}
    \label{fig:architect}
\end{figure*}

Figure \ref{fig:rt} displays DSAC's lifetime warranty according to the number of counters. In the case of 20 counters, the required time of R(\textit{t}) to be 0.999 is 9 days. This implies that DSAC can detect all the aggressor rows of double-sided uniform weight pattern for up to 9 days with a 1,000 parts per million (ppm) error rate. Note that 35 counters can guarantee 1 ppm error rate for 10 years. Therefore, DSAC can achieve a near-complete detection in low-area cost.

\section{Evaluation}\label{sec:6}
\subsection{Architecture of DSAC}\label{sec:6-1}
Figure \ref{fig:architect} illustrates the architecture of DSAC equipped with Time-Weighted Counter. TRR module comprises TRR module controller and TRR row detector.

\noindent\textbf{TRR Module Controller.} TRR Module Controller generates control signals for TRR row detector. 

{\footnotesize MAX\_OR\_MIN} is used to search for the maximum count row or the minimum count row for every {\footnotesize ACTIVE} which indicates active command. {\footnotesize ACTIVE} comes with {\footnotesize ACTIVE\_ROW\_[0:15]}, which indicates each bit of the row address for 64K rows. When Max. or Min. Scheduler receives {\footnotesize ACTIVE}, {\footnotesize MAX\_OR\_MIN} is low, and TRR Row Detector searches for the minimum count row to store {\footnotesize ACTIVE\_ROW\_[0:15]} into a count table. However, {\footnotesize STOCHASTIC\_REPLACEMENT} can block this operation. If {\footnotesize STOCHASTIC\_REPLACEMENT} is low, then {\footnotesize FILTERED\_ROW\_[0:15]} is equal to {\footnotesize ACTIVE\_ROW\_[0:15]} and can be stored into a count table. If {\footnotesize STOCHASTIC\_REPLACEMENT} is high, then {\footnotesize FILTERED\_ROW\_[0:15]} is filtered out. Once the minimum count row search is completed, {\footnotesize MAX\_OR\_MIN} becomes high, and TRR Row Detector searches for the maximum count row. 

To generate {\footnotesize STOCHASTIC\_REPLACEMENT}, which can filter out decoy-rows, Seed Mixer, LFSR, Min. Count Register, and Probability LUT are implemented. Seed Mixer receives {\footnotesize PUF}, which leverages DRAM's Physical Unclonable Function, so that its output {\footnotesize SEED\_MIXER\_[0:19]} can be unique to each DRAM. {\footnotesize SEED\_MIXER\_[0:19]} is updated for every tREFW using {\footnotesize ALL\_CELL\_REF\_DONE}, which indicates all cells are refreshed so that LFSR's {\footnotesize PRBS\_[0:19]} cannot be readily deciphered. LFSR updates its output {\footnotesize PRBS\_[0:19]} for every active command using {\footnotesize ACTIVE} to leverage probability for every {\footnotesize ACTIVE\_ROW\_[0:15]}. Min. Count Register receives all the row counts and outputs the minimum count {\footnotesize MIN\_CNT\_[0:13]} when {\footnotesize MAX\_OR\_MIN} is low. Probability LUT receives {\footnotesize MIN\_CNT\_[0:13]} and selects the corresponding probability generated by utilizing {\footnotesize PRBS\_[0:19]}. 20 bits are required to cover 2,095K MAC$_{\text{tREFW}}$. 

{\footnotesize PLUS\_OR\_MINUS} is used to calculate victim rows for every {\footnotesize TRR\_FLAG}, which indicates a refresh command that reaches TRR$_{\text{TH}}$. When Plus or Minus Scheduler receives {\footnotesize TRR\_FLAG}, {\footnotesize PLUS\_OR\_MINUS} is low, and Victim Row Cal. in TRR Row Detector calculates RH$-$\textit{x}. Once RH$-$\textit{x} calculation is completed, {\footnotesize PLUS\_OR\_MINUS} becomes high, and Victim Row Cal. in TRR Row Detector calculates RH$+$\textit{x}. Note that \textit{x} is non-negative integers to mitigate $\pm$\textit{x} rows adjacent to the aggressor row.

{\footnotesize COUNTER\_WEIGHT} is used to mitigate RowBleed and it can increment count value for a row that is activated longer than tRASmin. Note that a floating-point counter is not used for area reduction, and therefore the logarithm function of Equality \ref{eq2} is rounded up to the nearest integer.

\noindent\textbf{TRR Row Detector.} TRR Row Detector detects aggressor row {\footnotesize RH\_[0:15]} and outputs victim rows {\footnotesize VICTIM\_[0:15]}. The mechanism of TRR Row Detector is based on a single-elimination tournament where the loser of each match-up is eliminated from the tournament. Therefore, {\footnotesize RH\_[0:15]} is the winner of the final match-up. Note that the number of counters is scalable.

When {\footnotesize MAX\_OR\_MIN} is low, Comparator\#0(1)'s output {\footnotesize MAX\_OR\_MIN\#0(1)} selects the count value that is less between two Row Counters in Count MUX\#0(1). {\footnotesize MAX\_OR\_MIN\#0} and {\footnotesize MAX\_OR\_MIN\#1} are decoded to {\footnotesize PNT\_REG\#0/1/2/3}. {\footnotesize PNT\_REG\#0/1/2/3} is used to control Row Register and corresponding Row Counter. Since {\footnotesize MAX\_OR\_MIN} is low, the pointed Row Register replaces the stored row with a new row and corresponding Row Counter increments its count. If all the count values are the same, then the low index Row Register has a priority of replacement. 

When {\footnotesize MAX\_OR\_MIN} is high, Comparator\#0(1)'s output {\footnotesize MAX\_OR\_MIN\#0(1)} selects the greater count value between two Row Counters in Count MUX\#0(1). {\footnotesize MAX\_OR\_MIN\#0} and {\footnotesize MAX\_OR\_MIN\#1} are decoded to {\footnotesize PNT\_REG\#0/1/2/3}. {\footnotesize PNT\_REG\#0/1/2/3} is used to control Row Register and corresponding Row Counter. Since {\footnotesize MAX\_OR\_MIN} is high, the pointed Row Register sends its row to RowHammer Register. RowHammer Register outputs {\footnotesize RH\_[0:15]} when {\footnotesize MAX\_OR\_MIN\#2} is high. If all the count values are the same, then RowHammer Register selects the row from the high index Row Register to consider temporal locality.

When {\footnotesize ACTIVE} is high and tRAS is longer than tRASmin, Time-Weighted Counter's output {\footnotesize COUNTER\_WEIGHT} increments the corresponding row's count value. For example, if $\alpha$ is 1 and tRAS is $2\times\text{tRASmin}$, {\footnotesize COUNTER\_WEIGHT} becomes 1, and the corresponding row's count value becomes 2 (1 for normal counting and 1 for Time-Weighted Counting).

Row Counter is reset once TRR is performed. Note that if all the count values are 0, then no TRR is performed to save power consumption and to avoid TRR-induced bit-flip. Note that the number of counters is scalable.

\begin{table}[h]
\vspace{-1ex}
\centering
\caption{Real Chip Area Requirements}
\vspace{-1ex}
\label{tab:dsac}
    \begin{adjustbox}{width=\columnwidth}
        \begin{threeparttable}
            \begin{tabular}{lcc}
            \hlinewd{1.2pt}
            {} & \multicolumn{2}{c}{Area} \\ 
            {\multirow{-2}{*}{Module (Description)}} & {$\text{um}^{2}$} & {\% DRAM} \\ \hline
            {1 Chip (Major DRAM Manufacturer's 10nm-class 8Gb/Ch. LPDDR4)} & {32,510,639} & {100.000} \\ 
            {Row Register (16 Bits for 64K Rows of 8Gb/Ch. LPDDR4)} & {251} & {0.001} \\
            {Row Counter (14 Bits for 20K RH$_{\text{TH}}$ \cite{frigo2020trrespass})} & {162} & {0.000} \\
            {Comparator (14 Bits for Comparing Each Bit of 2 CNTs)} & {166} & {0.001} \\
            {2-to-1 MUX (14 MUXs for Selecting Each Bit of 2 CNTs)} & {125} & {0.000} \\
            {Decoder (Pointing Max. Cnt or Min. Cnt Row REG)} & {536} & {0.002} \\
            {RowHammer Register (16 Bits for 16Bit Row REG)} & {251} & {0.001} \\
            {Victim Row Cal. (Full Adder for Calculating Victim Row Addr.)} & {388} & {0.001} \\
            {Min. Count Register (14 Bits for 14Bit Row CNT)} & {219} & {0.001} \\
            {PRNG (20Bit LFSR for Uniform Distribution \& Seed Mixer)} & {463} & {0.001} \\
            {Probability LUT (Selecting the Corresponding Prob.)} & {14,810} & {0.046} \\
            {Time-Weighted Counter (Oscillator with Flip-Flops)} & {275} & {0.001} \\
            \rowcolor{hlg} {4 Count Table DSAC with Time-Weighted Counter for 8 Banks\tnote{*}} & {154,739} & {0.476} \\ \hlinewd{1.2pt}
            \end{tabular}
            \begin{tablenotes}
                \item[*] $[4\times(\text{REG} + \text{CNT}) + 3\times(\text{CMP}) + 2\times(\text{MUX}) + 1\times(\text{Decoder} + \text{RowHammer REG} + \text{Victim Row Cal.} + \text{Min. Cnt. REG} + \text{PRNG} + \text{LUT}) + 1\times(\text{Time-Weighted Counter})]\times8$
            \end{tablenotes}
        \end{threeparttable}
    \end{adjustbox}
\end{table}

Table \ref{tab:dsac} shows the required area for each module of DSAC based on real chip implementation. The bit-length of Row Register is determined by the number of rows per bank. For example, an 8Gb per channel LPDDR4 device that can have 64K rows per bank requires 16 bits to convert 16 bits to 64K ($2^{16}$). The bit-length of Row Counter is determined by RH$_{\text{TH}}$. For example, to count 16K ($2^{14}$) to consider double-sided hammer for 20K RH$_{\text{TH}}$ \cite{frigo2020trrespass}, 14 bits are required.

It is worth noting that if DRAM can implement a number of counters in the detection algorithm equal to the number of rows per bank, then all row activations can be precisely counted, allowing DRAM to effortlessly detect RowHammer. According to the baseline parameters in Table \ref{tab:base}, if DRAM reserves 64K counters per bank, there can be no errors in counts, and the TRR algorithm can accurately identify RowHammer attacks. However, implementing this would require DRAM to have 512K ($64\text{K}\times8$) counters for all banks, which is equivalent to the area of seven DRAM chips (512K Row Register$+$512K Row Counter).

\subsection{Maximum Disturbance}
This paper introduces a RowHammer protection index named Maximum Disturbance, which measures the maximum accumulated number of row activations within tREFW. For example, if a frequently accessed row is not TRRed within tREFW, then the accumulated number of accesses is recorded. The recorded number that exceeds RH$_{\text{TH}}$ indicates that TRR algorithm fails to protect DRAM against RowHammer attack.

Other TRR algorithms \cite{lee2019twice, park2020graphene, kim2014architectural, kim2014flipping, son2017making, you2019mrloc} are configured using the baseline parameters in Table \ref{tab:base}. Note that because DSAC does not require any external operation outside DRAM, this paper does not evaluate its impact on system performance.

\noindent\textbf{Malicious Attack Patterns.} The Maximum Disturbance of each TRR algorithm strongly depends on access pattern. This paper injects infamous RowHammer attack patterns called TRRespass \cite{frigo2020trrespass} and random access. In order to synthesize malicious RowHammer attacks, a double-sided uniform weight is adopted for both attack patterns. Note that double-sided denotes a victim row is positioned in between two aggressor rows \cite{frigo2020trrespass} and uniform weight denotes that all the incoming rows have uniformly distributed weights. The double-sided uniform weight is the worst pattern for DSAC as discussed in Section \ref{sec:5}. Table \ref{tab:attack} summarizes the injected malicious patterns. 

\begin{table}[h]
\vspace{-1ex}
\centering
\caption{Injected Malicious RowHammer Attacks}
\vspace{-1ex}
\label{tab:attack}
    \begin{adjustbox}{width=\columnwidth}
        \begin{tabular}{cccccc}
        \hlinewd{1.2pt}
        {Pattern} & {\# of Act.} & {\# of Rows} & {Row Sequence} & {Side} & {Weight} \\ \hline
        {TRRespass} & {MAC$_{\text{tREFI}}$} & {1$\sim$MAC$_{\text{tREFI}}$} & {Round-Robin} & {Double} & {Uniform} \\
        {Random} & {MAC$_{\text{tREFI}}$} & {1$\sim$MAC$_{\text{tREFI}}$} & {Random} & {Double} & {Uniform} \\ \hlinewd{1.2pt}
        \end{tabular}
    \end{adjustbox}
\end{table}

For TRRespass with 1 row, 1 row repeatedly accesses $\frac{255}{1}$ times in tREFI. For TRRespass with 100 rows, each of the 100 rows accesses $\frac{255}{100}$ times in tREFI in round-robin sequence. For TRRespass with 255 rows, each of the 255 rows accesses $\frac{255}{255}$ times in tREFI in round-robin sequence. 

For random access with 1 row, 1 row repeatedly accesses $\frac{255}{1}$ times in tREFI. For random access with 100 rows, each of the 100 rows accesses $\frac{255}{100}$ times in tREFI in random sequence. For random access with 255 rows, each of the 255 rows accesses $\frac{255}{255}$ times in tREFI in random sequence.

\begin{figure}[h]
    \centering
    \includegraphics[width=\columnwidth]{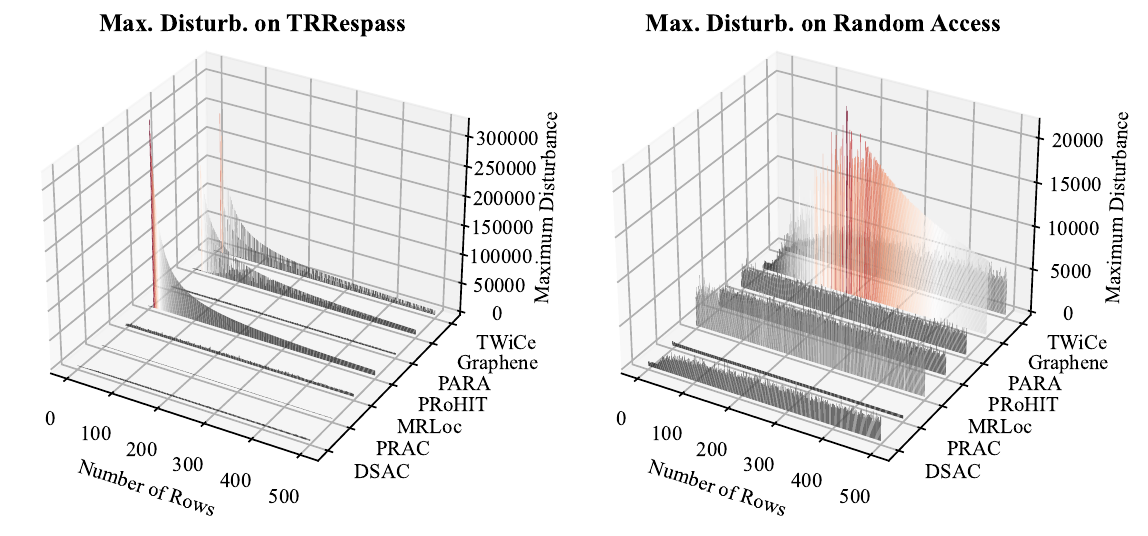}
    \vspace{-2em}
    \caption{Maximum Disturbance on Malicious Workloads}
    \label{fig:mali20}
\end{figure}

Figure \ref{fig:mali20} presents the results of the Maximum Disturbance experiment, with each TRR algorithm configured with 20 counters. The data confirm that DSAC's Maximum Disturbance increases when the number of rows becomes greater than the number of counters, 20 in this experiment, as discussed in Section \ref{sec:5}. Note that PRAC \cite{william2020per} represents per-row activation count, an ideal RowHammer detection algorithm capable of precisely tracking the activation count for all rows within a designated DRAM cell area.

For a double-sided attack to cause a bit-flip, aggressor rows need to reach a Maximum Disturbance of at least $\frac{\text{RH}_\text{TH}}{2}$, which is equal to 10K according to Table \ref{tab:base}. Since MAC$_{\text{tREFW}}$ is 2,095K, one of the 200 aggressor rows can reach 10K and become saturated. The data confirm that DSAC's Maximum Disturbance is saturated at around 200 rows for both TRRespass and random access pattern.

\begin{figure}[h]
    \centering
    \includegraphics[width=\columnwidth]{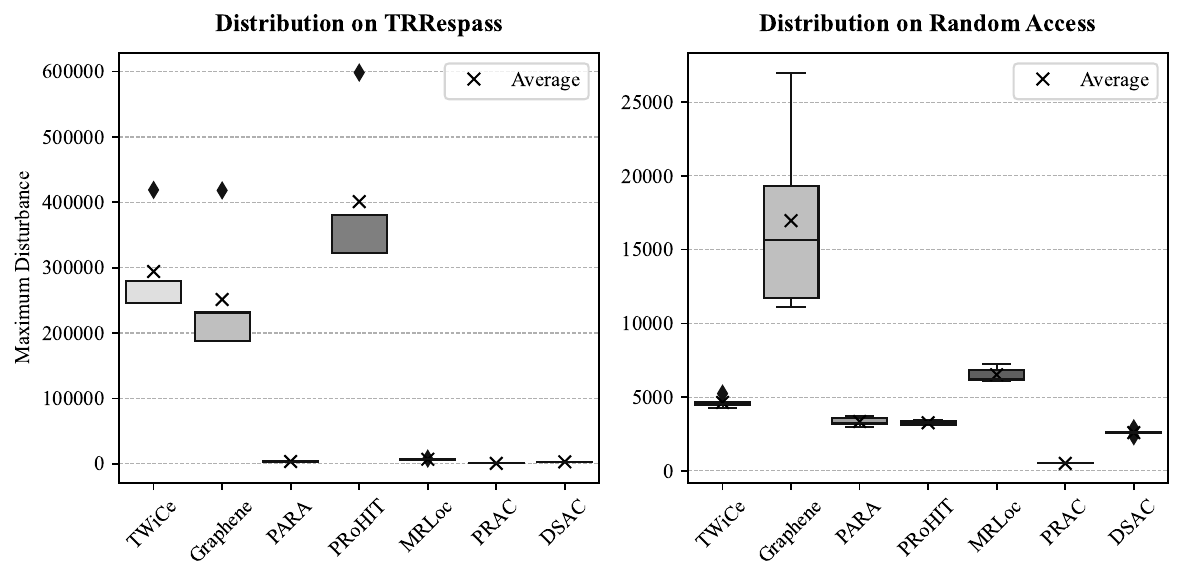}
    \vspace{-2em}
    \caption{Distribution of Maximum Disturbance on Malicious Workloads for Each TRR Algorithm Using 8 to 20 Counters}
    \label{fig:mali8_20}
\end{figure}

Figure \ref{fig:mali8_20} presents the results of the Maximum Disturbance experiment, elucidating the impact of the number of counters on the performance of each TRR algorithm. The range of counters spans from 8 to 20, allowing for an examination of the relationship between the number of counters and Maximum Disturbance. Furthermore, the number of rows ranges from 1 to 100, providing a comprehensive analysis of the effect of varying the number of rows on Maximum Disturbance.

\begin{figure}[h]
    \centering
    \includegraphics[width=0.6\columnwidth]{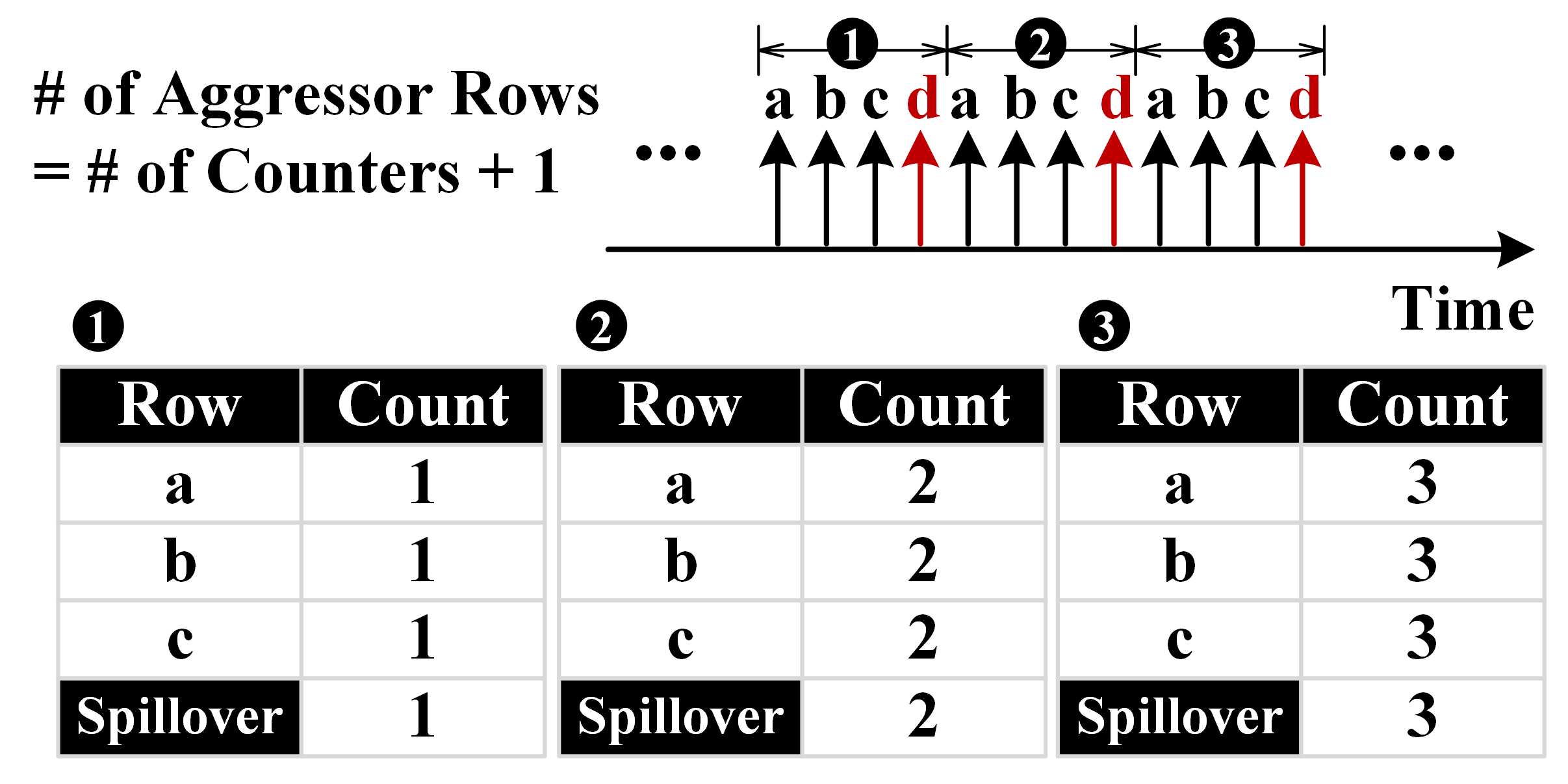}
    \vspace{-1em}
    \caption{Detection Failure of Low Area Cost Graphene}
    \label{fig:graphene}
\end{figure}

Graphene \cite{park2020graphene} exhibits a relatively high Maximum Disturbance, as it cannot filter out decoy-rows with fewer than 418 counters ($\text{the required number of counters}=\frac{\text{MAC}_\text{tREFWe}}{(\text{RH}_\text{TH}/4)+1}-1$), despite using the state-of-the-art counter-based data streaming algorithm called Misra and Gries algorithm \cite{misra1982finding}. Figure \ref{fig:graphene} illustrates the detection failure of low area cost Graphene. The number of aggressor rows is the number of counters$+$1 and the aggressor rows sequentially come in. Since decoy-row(d) cannot be inserted into a count table, the neighboring victim rows of decoy-row(d) cannot be TRRed.

\begin{table}[h]
\vspace{-1ex}
\centering
\caption{System Configuration for Benchmark Simulation}
\vspace{-1ex}
\label{tab:bench}
{\tiny
    \begin{adjustbox}{width=0.8\columnwidth}
        \begin{tabular}{
            >{}c 
            >{}c |c} 
            \hlinewd{1.2pt} 
            \multicolumn{2}{c|}{Parameter}                                & Configuration \\ \hline
            \multicolumn{2}{c|}{Number of Cores}                          & 16                                  \\ \hline
            \multicolumn{2}{c|}{Clock Frequency}                          & 2.5 GHz                             \\ \hline
            \multicolumn{1}{c|}{}                                         & L1D & 8-Way / 32 KiB per Core       \\ \cline{2-3} 
            \multicolumn{1}{c|}{}                                         & L1I & 8-Way / 32 KiB per Core       \\ \cline{2-3} 
            \multicolumn{1}{c|}{}                                         & L2  & 16-Way / 1 MiB per Core       \\ \cline{2-3} 
            \multicolumn{1}{c|}{\multirow{-4}{*}{Associativity / Size}}   & L3  & 11-Way / 22 MiB               \\ \hline
            \multicolumn{1}{c|}{}                                         & L1D & 4 Cycles                      \\ \cline{2-3} 
            \multicolumn{1}{c|}{}                                         & L1I & 4 Cycles                      \\ \cline{2-3} 
            \multicolumn{1}{c|}{}                                         & L2  & 10 Cycles                     \\ \cline{2-3} 
            \multicolumn{1}{c|}{\multirow{-4}{*}{Latency}}                & L3  & 46 Cycles                     \\ \hline
            \multicolumn{2}{c|}{Replacement Policy}                       & LRU                                 \\ \hline
            \multicolumn{2}{c|}{Main Memory Page Policy}                  & Closed                              \\ \hline
            \multicolumn{2}{c|}{Main Memory Data Rate}                    & 2933 Mbps                           \\ \hlinewd{1.2pt} 
        \end{tabular}
    \end{adjustbox}
}
\end{table}

\noindent\textbf{Benchmark Simulation.} In addition to evaluating the malicious RowHammer attack patterns, this paper measures Maximum Disturbance on workloads from the SPEC CPU 2017 benchmark suite \cite{speccpu2017}. This paper uses 17 rate-workloads, and for each workload, a 128ms snippet from a SimPoint \cite{hamerly2005simpoint} region consisting of 500 million instructions is used. Each experiment for each workload consists of 16 symmetrical processes simultaneously running identical workloads from an identical SimPoint region.

For the experiment, ZSim \cite{sanchez2013zsim}, an execution-driven system simulator based on the Intel Pin \cite{luk2005pin} instrumentation tool, is used. Similar to DRAMSim3 \cite{li2020dramsim3}, the simulator models the behavior of the memory controller, including read and write queues, address decoder, and DRAM command generation, as well as the bank-level DRAM device. The system and DRAM are configured as described in Table \ref{tab:bench}.

\begin{table*}[t]
\vspace{-1ex}
\centering
\caption{Summary of Maximum Disturbance on Malicious Workloads and SPEC CPU 2017 Benchmarks}
\vspace{-1ex}
\label{tab:sum}
    \begin{adjustbox}{width=\textwidth}
        \begin{threeparttable}
        \begin{tabular}{cccccccccc}
        \hlinewd{1.2pt}
        {Attack Pattern} & {Disturbance} & {Regular Refresh\tnote{*}} & {TWiCe \cite{lee2019twice}} & {Graphene \cite{park2020graphene}} & {PARA \cite{kim2014flipping}} & {PRoHIT \cite{son2017making}} & {MRLoc \cite{you2019mrloc}} & {PRAC \cite{william2020per}\tnote{**}} & \cellcolor{hlg}{\textbf{DSAC}} \\ \hline
        \multirow{3}{*}{TRRespass \cite{frigo2020trrespass}} & \cellcolor{lightgray1}{Maximum} & \cellcolor{lightgray1}{2,145,280} & \cellcolor{lightgray1}{419,000} & \cellcolor{lightgray1}{418,184} & \cellcolor{lightgray1}{3,532} & \cellcolor{lightgray1}{598,572} & \cellcolor{lightgray1}{7,784} & \cellcolor{lightgray1}{510} & \cellcolor{hlg}{3,138} \\
        {} & {Average} & {N/A} & {294,121} & {251,292} & {3,290} & {401,002} & {6,786} & {-} & \cellcolor{hlg}{2,780} \\
        {} & {Std. Dev.} & {N/A} & {71,717} & {95,764} & {202} & {114,266} & {616} & {-} & \cellcolor{hlg}{310} \\ 
        \multirow{3}{*}{Random Access} & \cellcolor{lightgray1}{Maximum} & \cellcolor{lightgray1}{2,145,280} & \cellcolor{lightgray1}{5,239} & \cellcolor{lightgray1}{27,006} & \cellcolor{lightgray1}{3,693} & \cellcolor{lightgray1}{3,448} & \cellcolor{lightgray1}{7,254} & \cellcolor{lightgray1}{510} & \cellcolor{hlg}{2,882} \\
        {} & {Average} & {N/A} & {4,638} & {16,967} & {3,349} & {3,260} & {6,518} & {-} & \cellcolor{hlg}{2,594} \\
        {} & {Std. Dev.} & {N/A} & {374} & {6,515} & {291} & {173} & {503} & {-} & \cellcolor{hlg}{192} \\ 
        \multirow{3}{*}{\begin{tabular}[c]{@{}c@{}}SPEC CPU 2017\\Benchmarks \cite{speccpu2017}\end{tabular}} & \cellcolor{lightgray1}{Maximum} & \cellcolor{lightgray1}{2,083} & \cellcolor{lightgray1}{1,377} & \cellcolor{lightgray1}{2,013} & \cellcolor{lightgray1}{1,527} & \cellcolor{lightgray1}{1,297} & \cellcolor{lightgray1}{1,531} & \cellcolor{lightgray1}{232} & \cellcolor{hlg}{832} \\
        {} & {Average} & {N/A} & {658} & {1,095} & {865} & {555} & {794} & {-} & \cellcolor{hlg}{482} \\
        {} & {Std. Dev.} & {N/A} & {409} & {696} & {503} & {370} & {499} & {-} & \cellcolor{hlg}{287} \\ \hlinewd{1.2pt}    
        \end{tabular}
            \begin{tablenotes}[flushleft]
            \item[*] Regular Refresh denotes a period refresh operation performed upon a periodic refresh command issued by memory controller, without involving TRR.
            \item[**] In this experiment, any row with Maximum Disturbance on every second refresh command is TRRed.
            \end{tablenotes}
        \end{threeparttable}
    \end{adjustbox}
\end{table*}

\begin{figure}[h]
    \centering
    \includegraphics[width=\columnwidth]{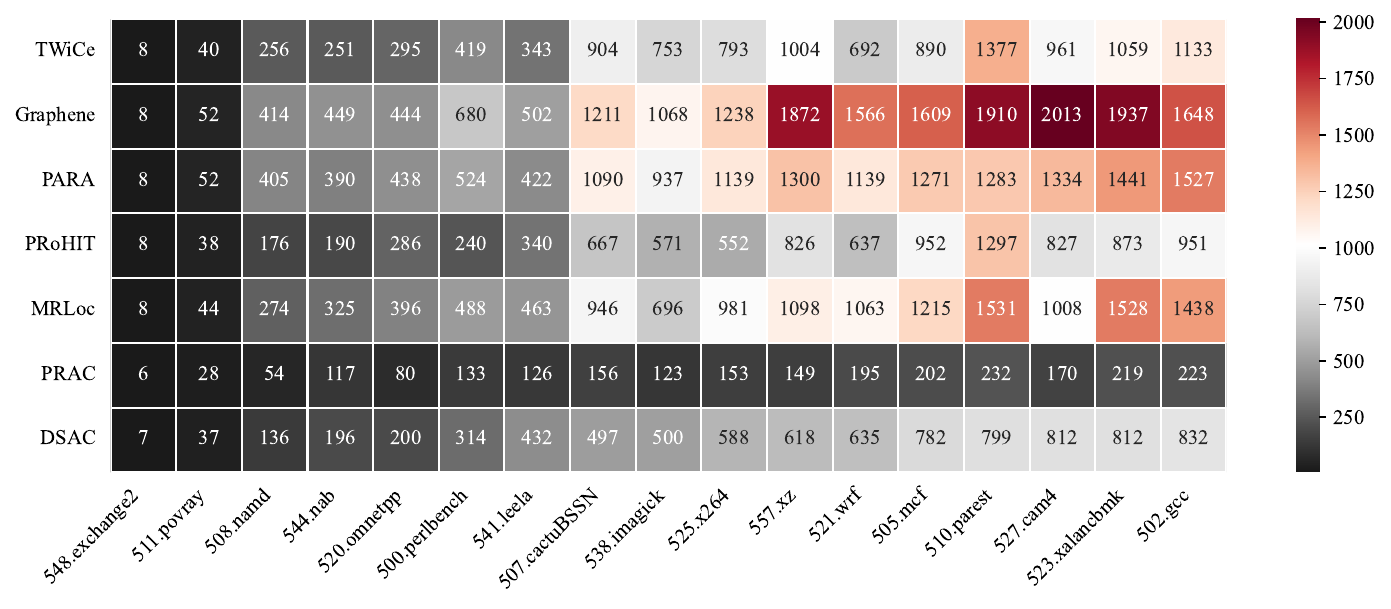}
    \vspace{-2em}
    \caption{Maximum Disturbance on SPEC CPU 2017 Benchmarks}
    \label{fig:heatmap}
\end{figure}

Figure \ref{fig:heatmap} presents the results of Maximum Disturbance experiment on the benchmarks, with each TRR algorithm configured with 20 counters. As the workloads do not involve adversarial RowHammer attacks, the Maximum Disturbances observed for all the workloads are lower than the Maximum Disturbances observed for malicious workloads. The data demonstrate that DSAC achieves the lowest Maximum Disturbance. Note that PRA \cite{kim2014architectural}, PARA \cite{kim2014flipping}, and PRoHIT \cite{son2017making} are susceptible to adversarial low locality patterns because they cannot filter out decoy-rows due to their constant probability.

The result data are summarized in Table \ref{tab:sum}. The data demonstrate that DSAC can achieve 133x lower Maximum Disturbance than the state-of-the-art counter-based algorithm named Graphene \cite{park2020graphene}. Moreover, the data demonstrate that DSAC exhibits the lowest average disturbance, indicating its robustness against various attack patterns. Additionally, DSAC's Maximum Disturbance is closest to an ideal detection algorithm named PRAC \cite{william2020per}.

\begin{table}[h]
\vspace{-1ex}
\centering
\caption{Comparison of TRR Algorithms}
\vspace{-1ex}
\label{tab:comp}
    \begin{adjustbox}{width=1\columnwidth}
        \begin{threeparttable}
        \begin{tabular}{lccccc}
        \hlinewd{1.2pt}   
        {Proposal} & {\begin{tabular}[c]{@{}c@{}}RowHammer\\Detection\end{tabular}} & {\begin{tabular}[c]{@{}c@{}}Decoy-Rows\\Filtering\end{tabular}} & {\begin{tabular}[c]{@{}c@{}}System\\Overhead-Free\end{tabular}} & {\begin{tabular}[c]{@{}c@{}}Scalability\tnote{*}\\for RH$_{\text{TH}}$\end{tabular}} & {\begin{tabular}[c]{@{}c@{}}RowBleed\\Mitigation\end{tabular}} \\ \hline
        CRA \cite{kim2014architectural}& {Deterministic} &  \ding{55} & \ding{55} & \ding{51} & \ding{55} \\ 
        CBT \cite{seyedzadeh2016counter}& {Deterministic} & \ding{55} & \ding{51} & \ding{55} & \ding{55} \\  
        CAT-TWO \cite{kang2020cat}& {Deterministic} &       \ding{55} & \ding{51} & \ding{55} & \ding{55} \\  
        TWiCe \cite{lee2019twice}& {Deterministic} &        \ding{55} & \ding{51} & \ding{55} & \ding{55} \\   
        Graphene \cite{park2020graphene}& {Deterministic} & \ding{55} & \ding{55} & \ding{51} & \ding{55} \\
        PRAC \cite{william2020per}& {Deterministic} &       \ding{55} & \ding{55} & \ding{51} & \ding{55} \\
        PRA \cite{kim2014architectural}& {Probabilistic} &  \ding{55} & \ding{55} & \ding{51} & \ding{55} \\  
        PARA \cite{kim2014flipping}& {Probabilistic} &      \ding{55} & \ding{55} & \ding{51} & \ding{55} \\  
        PRoHIT \cite{son2017making}& {Probabilistic} &      \ding{55} & \ding{51} & \ding{55} & \ding{55} \\ 
        MRLoc \cite{you2019mrloc}& {Probabilistic} &        \ding{55} & \ding{51} & \ding{55} & \ding{55} \\
        \rowcolor{hlg} \textbf{DSAC} & {Stochastic} &       \ding{51} & \ding{51} & \ding{51} & \ding{51} \\ \hlinewd{1.2pt}
        \end{tabular}
        \begin{tablenotes}[flushleft]
            \item[*] TRR algorithm is scalable if it can practically mitigate RowHammer for different RH$_{\text{TH}}$ by scaling their parameters such as the number of counters or TRR$_{\text{TH}}$.
        \end{tablenotes}
        \end{threeparttable}
    \end{adjustbox}
\end{table}

Table \ref{tab:comp} represents a comparison of TRR algorithms. DSAC is the first work that (1) countermeasures both RowBleed and RowHammer, (2) filters out decoy-rows statistically. Furthermore, DSAC requires (3) no system performance degradation since the operation of DSAC abides by memory standard specifications \cite{specification2014mjedec, specification2019jedec, specification2014jedec, specification2020jedec, specification2021gjedec, specification2021hjedec}.

However, PARA \cite{kim2014flipping} can be implemented with a comparatively smaller area, as it does not require counters, comparators, and multiplexers, which are necessary components in deterministic detection algorithms. According to Table \ref{tab:dsac}, PARA \cite{kim2014flipping} necessitates an area of 25,176$\text{um}^{2}$, calculated as $[4 \times \text{REG} + 1 \times (\text{Decoder} + \text{RowHammer REG} + \text{Victim Row Cal.} + \text{PRNG} + \text{LUT}) + 1 \times (\text{Time-Weighted Counter})] \times 8$ with a smaller Probability LUT. This area requirement is $-$84\% smaller than that of DSAC. Thus, if the area cost is the primary consideration, PARA \cite{kim2014flipping} can be chosen, given that the Maximum Disturbance is not significantly higher compared to DSAC.

Given that implementing per-row activation count with logic gates necessitates a considerable area, as discussed in Section \ref{sec:6-1}, PRAC \cite{william2020per} conducts per-row activation count within a designated DRAM cell area. When any row exceeds a predefined TRR threshold, it is queued in the RowHammer Register. Subsequently, TRR is executed upon a refresh command or a refresh management (RFM) command. Despite PRAC \cite{william2020per} being an ideal RowHammer detection algorithm, RowHammer can still occur if TRR is not performed when necessary. Therefore, defining TRR threshold and selecting a TRR candidate algorithm in the RowHammer Register, such as first-in-first-out or random selection, is crucial. In conclusion, PRAC \cite{william2020per} necessitates modifications to memory standard specifications, as well as to DRAM and system configurations, due to its requirement for read-modify-write operation to read the previous activation count and update the current activation count for every row activation.

\section{Conclusion}
This paper presents a thorough exploration of bit-flips induced by row activation in DRAM, specifically addressing RowBleed and RowHammer phenomena. RowBleed arises from charge leakage in a victim row due to a neighboring aggressor row's prolonged activation, leading to a decrease in the transistor's threshold voltage. To alleviate this issue, this paper proposes Time-Weighted Counting, which assigns greater counter weights to rows that are activated for longer durations.

In contrast, RowHammer occurs when a victim row experiences electron injection due to frequent activation of a nearby aggressor row. Extended RowHammer, the phenomenon where victim rows are two rows beyond aggressor rows, also results from electron injection due to repeated activation of a neighboring aggressor row. As a result, precise identification of aggressor rows is crucial. Therefore, this paper proposes RowHammer mitigation algorithm named DSAC, which can filter out decoy-rows by employing a replacement probability adjusted based on the old row count.

Additionally, this paper introduces a RowHammer protection metric called Maximum Disturbance, measuring the maximum accumulated number of row activations within an observation period. Experimental results demonstrate that DSAC outperforms the state-of-the-art counter-based algorithm, achieving a 133x lower Maximum Disturbance.

\bibliographystyle{IEEEtranS}
\bibliography{main}



\end{document}